\documentclass[a4paper,12pt]{article}
\usepackage{fullpage}
\usepackage[utf8]{inputenc}
\usepackage{authblk}
\usepackage[a4paper, margin=2.0cm]{geometry}
\usepackage{graphicx} 
\usepackage{float}
\usepackage{marvosym}
\usepackage[figuresright]{rotating}
\usepackage{hyperref}
\usepackage{listings}
\usepackage{amsmath}
\usepackage{pdflscape}
\usepackage{verbatim}
\usepackage{color}
\usepackage{wrapfig}
\usepackage{subcaption}
\usepackage{listings}
\usepackage{xcolor}
\usepackage{hyperref}
\usepackage{cite}
\usepackage{pdflscape}
\usepackage{booktabs}
\usepackage{float}
\usepackage{threeparttable}
\usepackage{apacite}
\usepackage{graphicx}
\usepackage{geometry}
\usepackage{pdflscape}
\usepackage{threeparttable}
\usepackage{geometry}
\usepackage{array}
\usepackage{hyperref}
\usepackage{authblk}
\usepackage{lineno}
\usepackage{setspace}
\begin{document}

\title{\vspace{-2.0cm}Physics-based deep learning reveals rising heating demand heightens air pollution in Norwegian cities}
\author[1,2]{Cong Cao}
\author[1,3]{Ramit Debnath\textsuperscript{\Letter}}
\author[1]{R. Michael Alvarez}
\affil[1]{California Institute of Technology, USA}
\affil[2]{Norwegian University of Science and Technology, Norway}
\affil[3]{University of Cambridge, UK}
\affil[\Letter]{corresponding author: rd545@cam.ac.uk}

\date{} 
\renewcommand\Affilfont{\small}

\maketitle
\section * {Abstract}
Policymakers frequently analyze air quality and climate change in isolation, disregarding their interactions. This study explores the influence of specific climate factors on air quality by contrasting a regression model with K-Means Clustering, Hierarchical Clustering, and Random Forest techniques. We employ Physics-based Deep Learning (PBDL) and Long Short-Term Memory (LSTM) to examine the air pollution predictions. Our analysis utilizes ten years (2009-2018) of daily traffic, weather, and air pollution data from three major cities in Norway. Findings from feature selection reveal a correlation between rising heating degree days and heightened air pollution levels, suggesting increased heating activities in Norway are a contributing factor to worsening air quality. PBDL demonstrates superior accuracy in air pollution predictions compared to LSTM. This paper contributes to the growing literature on PBDL methods for more accurate air pollution predictions using environmental variables, aiding policymakers in formulating effective data-driven climate policies.
\section*{Keywords}
Deep Learning, AI, Physics-Informed Neural Networks, Climate Change, Air Quality

\newpage

\section{Introduction}
Exposure to high levels of air pollution has significant negative health outcomes \cite{Currie2004}. Concerns related to this have encouraged policymakers to consider the effectiveness of transportation and air quality policies to eliminate the adverse impact on health \cite{CohenRossGBD2005} and \cite{CAO2024544}. However, this requires better forecasting and estimation of the link between traffic, weather, and air pollution. In practice, the scarcity of comprehensive, high-quality air pollution data presents a limitation for both research and policy development in many countries. At the same time, accurately quantifying the health impacts of pollution to propose optimal air quality (AQ) policy remains challenging, even for a country like Norway, where pollution levels are already relatively low.\par

Climate change and air pollution are deeply interconnected because the chemicals that lead to degradation in AQ are frequently co-emitted with greenhouse gases \cite{Orru_Ebi_Forsberg_2017}. This interdependency means that interventions to address one will generally cause changes in the other, which in turn, can substantially increase the compounded risks of climate impacts and AQ on the population \cite{PinhoHolgate_2023}. Several examples show that there are co-benefits of net zero policies like reducing emissions from energy, transport, space heating, and cooking on reducing CO2, NOx, and PM emissions. Therefore, it is important to study the interactive effects of climate change and AQ simultaneously to find appropriate mitigation strategies. 

However, data quality and methodological limitations remain a significant challenge. Traditional time-series-based statistical methods pose challenges in examining climate impact and AQ data together due to their sensitivity to the sample size, which makes them prone to overfitting \cite{Babyak2004}. Moreover, if there are many covariates, traditional statistical methods tend to manually weigh each covariate, making it hard to estimate results with greater accuracy \cite{Yang2023}. Newer Machine Learning (ML) and Artificial Intelligence (AI)-led approaches are showing tremendous promise in bridging the aforementioned gap and have been successful in providing high accuracy with classification and prediction tasks of interactive environmental impact \cite{Black2023}. For instance, Chau et al. (2022) \cite{Chau2022} showed that Long Short-Term Memory (LSTM) and Bidirectional Recurrent Neural Network (BiRNN) methods can be used in deep learning-based weather forecasting models to measure changes in AQ. LSTM has an independent memory unit; it can remember longer-term information for a long time and avoid the problem of vanishing gradients. Consequently, LSTM is particularly useful for time series data \cite{Yadav2020}.


However, LSTM requires more training data in order to learn effectively, is computationally intensive, and can be slow to train on large datasets \cite{you2019large}, making it not appropriate for weather or physical-process data like the fluid flow of the atmosphere. Nonetheless, there has been rapid development in solving partial differential equations representing physical processes, such as the fluid dynamics of air pollution. Recent work has used ML by training deep neural networks through the enforcement of physical laws, which improve model generalizability, explainability, and accuracy \cite{KarniadakisKevrekidis2021, wang2023scientific}. This approach is called Physics-Based Deep Learning (PBDL), which uses (noisy) data and mathematical models and implements them through neural networks or other kernel-based regression networks \cite{KarniadakisKevrekidis2021}. In this paper, we expand the current PBDL's capabilities to inform policy design and subsequent decision-making by testing its effectiveness in predicting the interaction effects of air pollution under local traffic and weather conditions in Norway. \par

Our aim in feature selection is to identify the variables that have the most significant impact on air pollution. Simultaneously, we compare the consistency and differences in classification results between traditional panel models and machine learning algorithms. This step primarily serves as an exploratory analysis. The reason for conducting this is to propose corresponding policy recommendations after identifying the most influential variables. We used three machine learning algorithms, namely K-Means (KM), Hierarchical clustering (HC), and Random Forest (RF), and compared the feature selection results with the traditional panel model. During the feature selection phase of building the panel model, we strive to surpass existing EU standards by integrating complementary meteorological variables critical for air pollution and incorporating traffic flow conditions.\par

Better forecasting performance is crucial for making effective policy recommendations. We employ physics-based deep learning algorithms and time series model-based deep learning algorithms to compare the predictive performance of these two-dimensional approaches on time series data. These two methods are  Physics-based Deep Learning (PBDL) and Long Short-Term Memory (LSTM). We aim to identify the optimal models suitable for predicting traffic, meteorological, and air pollution variables. We seek to address the following research questions: Through feature selection, can we determine which climate factors are most important for air quality? Can deep learning better predict air pollution based on local traffic and weather conditions? Finally, will policy changes affect the predictive accuracy of deep learning algorithms? By answering these questions, we highlight the importance of extreme weather conditions in assessing the effects of air pollution.\par

Our study makes two unique and timely contributions to the current literature. First, we identify the most critical influences on air quality and provide insights that contribute to the design of highly accurate, evidence-based climate policies. Second, we highlight the methodological applicability of PDBL in complex decision-making tasks.\par

We provide further details about PBDL in environmental decision-making applications based on the current literature and its contextualization for the Norwegian air pollution and climate change case in Section 2. We expand the methodological details in Section 3, followed by results in Section 4. Section 5 concludes a discusses the implications of the results.

\section{Background}
We began our exploration with Physics-informed Machine Learning in conjunction with the atmospheric and climatic conditions in Norway. The Machine Learning algorithms we utilize can scrutinize the physical characteristics present in environmental datasets. We also seek to ascertain whether changes in Norwegian policies have any impact on the accuracy of deep learning results.\par

\subsection{Physics-informed ML for environmental modeling}
Physics-informed ML integrates seamlessly data and mathematical physics models for scientific discovery, even in partially understood, uncertain, and high-dimensional contexts, and is a rapidly evolving field \cite{KarniadakisKevrekidis2021}. The State-Of-The-Art (SOTA) approaches involved kernel-based or neural-network-based regression methods for effective and meshless implementations of parameterized partial differential equation solutions, even with noisier datasets \cite{KarniadakisKevrekidis2021, hao2022physics}. The growing application portfolio of physics-informed ML ranges from earth system-level models to models of lower-level regional systems.

For instance, a recent study employed a Physics-Informed Neural Network (PINN) by integrating deep learning and a hydro-physical model to capture credible runoff predictions under the influence of climate change \cite{Nazari_9863669}. This study demonstrated by designing deep learning models informed by hydrological processes, the predictive accuracy of the model improved significantly.  Another study used PINN to predict air pollution and found that the PINN algorithm, combining atmospheric physical and chemical models with deep learning, better captures the spatial distribution characteristics of the atmosphere, thereby improving the predictive accuracy of the model \cite{zhang2020first}. This study emphasized the need to describe uncertainties in model predictions for increasing the reliability of the PINN algorithms. 

In 2023, a new physics-informed approach was introduced: the Physics-Informed LSTM (PT-LSTM) algorithm. It considers the mass conservation law of physics and was demonstrated to be robust in predicting the salinization of Belgian peer waterways. This approach exhibited superior performance compared to traditional machine learning methods \cite{BERTELS2023129354}.The study noted that future research could focus on modeling non-conservative pollutants \cite{BERTELS2023129354}. Air pollutants related to non-conservative pollutants include gaseous substances present in the atmosphere at concentrations that often cause harm to flora, fauna, or the environment.

In a novel approach, PINN has recently been applied in epidemiology. In a study aimed at understanding the time evolution of infectious diseases during the COVID-19 pandemic, researchers applied PINN to German COVID-19 transmission dynamics data. The study found that PINN can accurately predict the dynamics of virus transmission \cite{han2024approaching}. Similarly, researchers have used PINN to explore whether it can identify and predict virus transmission trends, as well as prediction accuracy, by comparing it with multiple traditional epidemiological models. The study employed data from COVID-19 reports across the United States and Italy. The results indicate that across different datasets, PINN provides more accurate predictions compared to the traditional epidemic models \cite{kharazmi2021identifiability}. A study on COVID-19 in Stockholm utilized the PINN framework to estimate epidemiological parameters and found that PINN can reliably identify the susceptible-exposed-infectious-recovered model without making any a priori assumptions about the shape of infectiousness, thus demonstrating the advantages of PINN in epidemiological dynamic prediction \cite{tronstad2022physics}, which can also be applied for air pollution transport problems.

Specific earth and environmental science applications of PINN include the estimation of subsurface properties such as rock permeability and porosity, from seismic data by coupling neural networks with full-waveform inversion, subsurface flow processes, and rock physics models \cite{li2020coupled}. Similarly, Zhu et al. \cite{zhu2021general} demonstrated that PINN is capable of solving a wide number of seismic inversion problems like velocity estimation, earthquake location, and fault rupture imaging. PINNs are also used for improving weather prediction of wind power forecasting \cite{mammedov2022weather}. Liu-Schiaffini et al. \cite{liuschiaffini2023tipping} have used a variant of PINN with a recurrent neural operator to forecast a climate tipping point in stratocumulus cloud cover which has direct implications for studying the effects of greenhouse gas concentration in global cloud covers. For a more comprehensive list of weather and climate modeling applications, see \cite{kashinath2021physics}.  

In air quality assessment, Li et al. (2023) \cite{li2023improving} demonstrated the use of physics-inspired deep graph learning that encodes fluid physics to capture spatial and temporal dynamic characteristics of air pollutants in China. Their approach showed consistently improved extrapolation accuracy by an average of 11–22\% compared to several baseline machine learning methods. The authors note that PINNs can further improve physically relevant results. Similarly, Yang et al. \cite{yang2023new} built a Physics-Informed Multi-Task Deep Neural Network (phyMTDNN) for the joint retrieval of six main air pollutants, i.e., PM2.5, PM10, SO2, NO2, CO, and O3 using ground-based air pollution data from China and satellite reflectance data from Landsat-8. 

\subsection{Norwegian air quality policies and climate impacts}

We use Norway as a case study to explore the dynamics of air pollution and climate change. Due to the Norwegian government's many years of high-quality air pollution monitoring and data collection programs. The data, obtained from air pollution detectors installed on highways, feature high frequency, few missing values, and overall high quality \cite{ALDRIN20052145}. 

The colder climate of Norway and its recent exposure to climate-induced heat waves and other extreme weather conditions provide an interesting case study to examine air pollution and climate change impacts.  Moreover, Norway is a global leader in sustainability policy-making \cite{Vestreng2014}. According to the well-known Arcadis Sustainable Cities Index, the capital Oslo was awarded the title of the first most sustainable city in the world in 2017 \cite{Pozdniakova2017}. However, despite this, concerns remain about the harmful effects of transport-related pollution in Norway under climate change, and our paper seeks to understand this complex relationship. Norway’s beautiful natural landscape serves as a real-world laboratory, providing insights and lessons for our research. We focus on Norway not only because of its geographical location but also because of Norway’s multi-faceted exploration of air quality policies, energy innovation, and sustainable development. We try to explore if the policy change will affect the deep learning prediction accuracy,   more details are in Appendix A.\par


In terms of climate impacts on air pollution in Norway, a recent study showed that wind direction has an important impact on air pollution concentrations in Oslo, and evaluates the impact of Norwegian air quality policies, arguing that the model needs to consider meteorological variables. They discussed the practical challenges of traffic control policies. To reduce traffic congestion and improve air quality, establishing low-emission zones and increasing parking fees are the most effective permanent measures while banning diesel vehicles is the most effective temporary traffic control  \cite{Santos}. 

Furthermore, a survey conducted in 2016 indicated two-thirds of Norwegians feel the impact of climate change in Norway. Norway has implemented various climate-related policies, including the introduction of a carbon dioxide tax in 1991, subsidies for electric vehicles, a prohibition on the use of oil for heating, etc  \cite{steentjes2017european}. Additionally, Norway has committed to and has successfully achieved a reduction of approximately four percent in greenhouse gas emissions by 2020 compared to 1990 levels as per Kyoto Protocol \cite{BREMER2020100236}. Therefore, the future of air quality and climate policies for Norway must be aligned to maintain its global leadership position, and our study provides a unique methodological basis for it. 

\section{Data and methods}
The broad methodological framework is contributing to the growing field of computational social sciences. We use the Norwegian administrative time-series daily data from 2009 to 2018 that includes traffic, multiple weather factors, and air pollution. We selected three populous cities in Norway (Oslo, Bergen, and Trondheim) for this study. We used recurrent neural network-based Long Short-Term Memory (LSTM) \cite{hua2019deep} and Physics-Based Deep Learning (PBDL) for air pollution prediction  \cite{KarniadakisKevrekidis2021}. 

Simultaneously, we employed three commonly used ML algorithms for feature selection, namely K-means clustering, hierarchical clustering, and random forest, aiming to compare them with traditional statistical models for robustness checks. In the study design, we considered policy changes as influential factors in model predictions. Therefore, we take into account the interactive and lagged effects of air pollution and weather factors, moving beyond the consideration of single factors. Figure 1 provides a comprehensive overview of this methodology. 

The dataset came from three sources: traffic, weather, and air pollution. Initially, we conducted feature selection and compared the results of the most important variable selection between the traditional panel model and the ML algorithms. Subsequently, we employed two deep-learning algorithms to compare prediction performance. Specifically, Long Short-Term Memory (LSTM) is a deep learning algorithm well-suited for time series data, capable of capturing temporal relationships in the data. A Physics-Based Deep Learning (PBDL) is an algorithm that combines physical modeling with deep learning technology \cite{raissi2019physics}\cite{kashinath2021physics}.

\begin{figure}[H]
		\centering
		\includegraphics[width=0.9\linewidth]{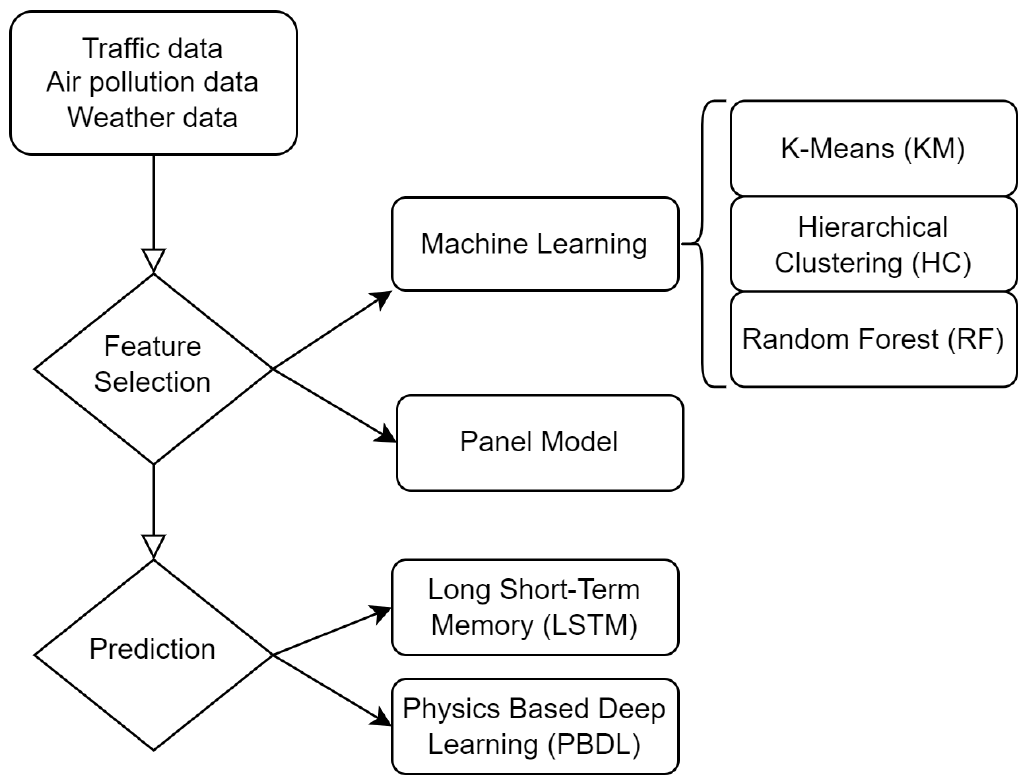}
		\caption{The methodological flow chart. Initially, we conducted feature engineering on the dataset to identify the primary climate factors influencing air pollution. Subsequently, we employed a panel regression model, along with three machine learning algorithms: K-means, Hierarchical clustering, and Random Forest, to explore and compare the feature selection outcomes of these models. Next, we evaluated the predictive performance of air pollution using PBDL and LSTM models. Finally, we integrated the feature selection and prediction outcomes to draw our conclusions.}
		\label{Oslo}
\end{figure}

\subsection{Data sources}

\subsubsection*{Air pollution}

The air pollution data comes from the Norwegian Institute for Air Research (NILU). We use daily air pollution data at the monitoring station level, including $PM_{2.5}$ and $NO_{x}$. The unit of both pollutants is represented in micrograms per cubic meter (µg/m³). 

We chose these two pollutants because they are known to have adverse health effects even at low concentrations. For example, 
at low concentrations of $NO_{x}$ can cause symptoms such as coughing, shortness of breath, nausea, tissue hypoxia accumulation of body fluids, and even death \cite{César2015}. $NO_{x}$ is unstable in sunlight and can produce ozone, which can damage human lung functions, such as increasing the incidence of asthma. $NO_{x}$ can also form adsorbent particles of matter through chemical reactions in the air, which will absorb microbial components such as bacteria in the air and become more harmful pollutants. Regarding $PM_{2.5}$, it can directly reach the lung tissue and cause pulmonary heart disease, aggravating asthma, and chronic bronchitis, especially for the elderly and children with weak constitution \cite{Xing2016}. 
 
\subsubsection*{Traffic data}

Motor vehicles are known to be one of the main sources of air pollution \cite{AGUDELOCASTANEDA201945}. Traffic volume data comes from the Norwegian Public Road Administration (SVV), which provides the traffic data on the road network, and is measured on-site from traffic registration points on municipal roads. We use daily data, which represents the number of motor vehicles passing through monitored traffic stops within 24 hours. All data are formulated and corrected following National Highway Administration standards, so data quality is guaranteed. The daily traffic flow we get is the sum of hourly traffic flow within 24 hours from 2009 to 2018.

We focus on the three large cities in Norway, Oslo, Bergen, and Trondheim. This reflects both data availability and that these are, naturally, the places with the highest concentrations of emissions. We selected monitoring stations that had both air pollution data and traffic flow data. Because individual sites often contain missing values, we did not treat each site as a cross-sectional observation but instead took the average value of the stations in each city. \par

\subsubsection*{Weather data}

We obtain daily measurements of weather data using monitoring stations from the Norwegian Meteorological Institute (MET). The data includes mean air temperature, heating degree days, vapour pressure, average mean wind speed, mean wind gust, mean relative humidity, and snow depth.
The proportion of missing values for the entire raw dataset is 7.89\%, thus, we use the “mice” package in R v4.3.2, to impute to get the complete dataset. The method we use in the R package is logic regression, and the missing daily observation is approximated by the other non-missing values with the logistic regression model.

We added time variables, including day, day of the week, month of the year, and year. The summary statistics table is illustrated in Table 1. We also took into account Heating Degree Days (HDD). HDD is the total number of degrees Celsius between the daily mean temperature and the threshold temperature, set at 17 degrees Celsius in Norway. Global warming is expected to decrease the count of HDD days, potentially affecting energy demand during winters \cite{spinoni2018changes},\cite{seljom2011modelling} .\par

\thispagestyle{empty}
\begin{table}[H]
    \centering
    \begin{threeparttable}
        \caption{Summary statistic table of the dataset}
        \label{tab:rotatedsamplelscape}
        \small
        \begin{tabular}{ccccccc}
             \hline
            \textrm{Variable} & \textrm{Abbreviation} & \textrm{Obs} & \textrm{Mean} & \textrm{Std.Dev.} & \textrm{Min} & \textrm{Max} \\
            \hline
            Day of the week & weekday & 14,606 & 4.000 & 2.000 & 1 & 7 \\
            Day of the month & day of M & 14,606 & 15.730 & 8.800 & 1 & 31 \\
            The month of the year & monthly & 14,606 & 6.553 & 3.440 & 1 & 12 \\
            Year & year & 14,606 & 2013.729 & 2.790 & 2009 & 2018 \\
            Day & day & 14,606 & 7496.277 & 4262.794 & 1 & 14606 \\
            Traffic volume (Passenger Car Unit) & TV & 14,606 & 1082.102 & 623.334 & 29504 & \\
            Nitrogen oxides ($\mu$g/$m^{3}$) & $NO_{x}$ & 14,606 & 80.398 & 69.020 & 0 & 361 \\
            Particulate matter ($\mu$g/m$^3$) & $PM_{2.5}$ & 14,606 & 10.388 & 7.3889 & 0 & 53 \\
            Mean air temperature (°C) & Tmean & 14,606 & 1801.186 & 1045.362 & -23 & 34 \\
            Heating degree days (17°C) (°C) & HDD & 14,606 & 30.126 & 70.637 & 1 & 308 \\
            Vapour pressure 24h (kPa) & VP & 14,606 & 27.953 & 53.399 & 0 & 182 \\
            Average mean wind speed 24h (m/s) & WS & 14,606 & 7.655 & 12.960 & 1 & 82 \\
            Mean wind gust 24h (m/s) & WG & 14,606 & 21.684 & 37.958 & 1 & 135 \\
            Mean relative humidity 24h (\%) & mean RH & 14,606 & 12.409 & 21.169 & 1 & 71 \\
            Snow depth 24h (cm) & SD & 14,606 & 3.348 & 9.072 & 0 & 67 \\
            Precipitation 24h (mm) & PP & 14,606 & 11.253 & 39.327 & 0 & 238 \\
           \hline
        \end{tabular}
        \begin{tablenotes}
            \item Note: Obs = Observations, Mean = Mean value, Std.Dev. = Standard Deviation.
        \end{tablenotes}
    \end{threeparttable}
\end{table}

\subsection{ML strategy}
 
\subsubsection*{K-Means (KM) }
K-Means is a popular clustering algorithm that groups data points based on their similarities. First, we decide how many groups we want. Then, we randomly choose K points from the data as centroids. Next, we calculate the distance from each data point to these centroids and assign each point to the closest centroid. At this point, each centroid has data points around it. We calculate a new centroid for each group of data points, creating a new K point. We repeat this process until the new centroids are very close to the old ones, indicating stable clustering results and converged data. \cite{wu2012cluster}.

\subsubsection*{Hierarchical clustering (HC) }\par

Hierarchical clustering algorithms form a hierarchical clustering tree by calculating the similarity between data points. The original data point is the bottom of the tree, and the top is the root node of a cluster. From the leaf node to the root node, it is completed by a merge algorithm. The combination is performed by calculating the similarity between the two data points. The process of calculating the similarity is to calculate the distance between the two data points. The smaller the distance, the higher the similarity \cite{nielsen2016hierarchical}. 

The process of calculating distance and combination is as follows: First, obtain an $N \times N$ matrix, where $X[i][j]$ refers to the distance between points $i$ and $j$. Set each data point to $d_i, i \in \{0,1,\ldots,N\}$.
By merging the data points with the smallest distance, a new combined data point $M$ is obtained, and the distance between the data points $M$ and $d_i$ is calculated. The calculation methods include Single Linkage, Complete Linkage, Average Linkage, and so on.

\subsubsection*{Long Short-Term Memory (LSTM)}\par
Time series data commonly exhibit prolonged dependencies, meaning that information from previous instances significantly influences predictions at the present instance. Traditional neural networks frequently falter in handling these prolonged dependencies, hindering the model's ability to grasp long-term memory. LSTM, however, effectively manages this challenge by regulating information flow through its gating mechanism, comprising forget gates, input gates, and output gates. This mechanism enables the model to selectively discard or retain particular information, thereby enhancing its capability to capture long-term dependencies within time series data.\cite{hochreiter1997long}.\par

\subsubsection*{Physics-Based Deep Learning (PBDL)}

Deep learning algorithms require extensive datasets for training, yet they often neglect underlying physical principles, leading to potential inaccuracies in prediction. In 2019, when the Brown University research team first proposed Physics-Informed Deep Learning (PIDL) or Physics-Based Deep Learning (PBDL), PIDL garnered significant attention \cite{Raissi2019}. Subsequently, several similar physics-based deep learning algorithms emerged, with applications in improving the accuracy of climate models by combining deep learning and computational fluid dynamics. PINN is currently one of the prominent models in the field of physics-based artificial intelligence.is to add the difference before and after the iteration of the physical partial differential equation to the loss function of the neural network. Therefore, when the algorithm is iteratively trained, not only the optimization of the loss function but also adheres to the laws of physics. 

Given that the data we utilize primarily pertains to temporal variations and lacks a distinct spatial dimension, it represents the fluctuations of different variables across various time points. Therefore, we opt for employing Ordinary Differential Equation (ODE) as the framework integrated into deep learning, specifically chosen to delineate the diffusion and chemical reactions of $\text{NOx}$ and $\text{PM2.5}$ \cite{alexandrov1997numerical}. We employ the ensuing ODE equation to portray alterations in air quality:

\begin{align}
\frac{d}{dt} \text{NOx}(t) &= f(\text{TV}, \text{Tmean}, \text{HDD}, \text{VP}, \text{WS}, \text{WG}, \text{meanRH}, \text{SD}, \text{PP}) \tag{1}
\end{align}

Wherein, $\text{NOx}(t)$ denotes the NOx concentration at a specific time point $t$, while\begin{align*}
f(&\text{TV}, T_{\text{mean}}, \text{HDD}, \text{VP}, \text{WS}, \text{WG}, \text{meanRH}, \text{SD}, \text{PP})
\end{align*}
 stands as a function of a series of variables encompassing Traffic Volume (TV), mean Temperature (Tmean), Heating Degree Day (HDD), Vapor Pressure (VP), mean Wind Speed (WS), mean Wind Gust (WG), mean Relative Humidity (meanRH), Snow Depth (SD), and Precipitation (PP). In the PBDL model, we shall employ ODE alongside neural networks to fit this equation, ensuring that the model grasps the fundamental physical principles elucidating the system's behavior. The same approach applies to PM2.5.\par

The primary distinction between the two deep learning algorithms lies in the calculation of the total loss. In PBDL, the total loss is the weighted sum of the initial condition loss, determined by comparing the predicted value to the real value, and the ordinary differential equation (ODE) loss, which measures the deviation between the predicted value and the differential equation loss satisfying the ODE definition \cite{Raissi2019}. The other factor that influences the prediction performance of deep learning is the adjustment of hyperparameters, such as network depth, learning rate design strategy, etc., data can also affect algorithm performance. Despite the identical data sample size and quality, $PM_{2.5}$ and $NO_{x}$ exhibit different data diversity. The diversity of data can help the algorithm better fit and predict the data set. 
 
\subsubsection*{Panel Model}
The primary models utilized for assessing climate change and air quality within the European Union are the ``Computer Model to Calculate Emissions from Road Traffic'' (COPERT)\cite{cao2024better} and the Greenhouse Gas and Air Pollution Interactions and Synergies (GAINS) Model\cite{amann2011cost}.\footnote{Available at \url{https://web.jrc.ec.europa.eu/policy-model-inventory/explore/models/model-gains/} and \url{https://web.jrc.ec.europa.eu/policy-model-inventory/explore/models/model-copert/}.}  The COPERT model is utilized for policy assessment in environmental research across European nations. Yet, although it considers factors like temperature and humidity, it lacks the incorporation of a broader range of meteorological variables influencing air quality. \par
On the other hand, the GAINS model is employed to analyze the effects of reducing air pollution and greenhouse gas emissions on health and the environment, aiming to fulfill specific environmental policy objectives. It integrates emission control policies across national boundaries and various sectors, serving as a crucial tool for the European Commission in policy analysis and air policy evaluation. However, while the GAINS model addresses the interplay between air pollution and climate change, it has limitations in sectors such as agriculture and energy, with no inclusion of the transportation domain. Our proposed model aims to surpass these established EU standards by incorporating additional meteorological variables pertinent to air pollution and integrating traffic flow conditions. We also employ quadratic and cubic functions to capture nonlinear relationships effectively.\par
Our empirical approach is Pollution = f (Traffic, Weather). We use pollutants as dependent variables, and weather and traffic as independent variables. We include quadratic and cubic terms for traffic and weather factors to form multivariate polynomial equations:\par

\begin{align}
P_{c,t} = & \alpha + \sum_{t=1}^T \sum_{c=1}^4 \beta_{\delta} T_{c,t} + \sum_{t=1}^T \sum_{j=1}^8 \zeta_j W_{t,j} \nonumber \\
& + \sum_{t=1}^T \sum_{c=1}^4 \beta_{\delta} T_{c,t}^2 + \sum_{t=1}^T \sum_{j=1}^8 \zeta_j W_{t,j}^2 \nonumber \\
& + \sum_{t=1}^T \sum_{c=1}^4 \beta_{\delta} T_{c,t}^3 + \sum_{t=1}^T \sum_{j=1}^8 \zeta_j W_{t,j}^3 + DW + DM + MY + \epsilon + \text{Dummy} \tag{2}
\end{align}
 \\

Where $P_{c,t}$  is air pollution, is the outcome of interest, including pollutant $NO_{x}$ or $PM_{2.5}$, t is time, from 1,2,3…,t, the unit is one day, c is a different city in Norway, from 1,2, …,4;$T_{c,t}$ is traffic volume, $W_{t,j}$  is weather, j is weather variables number, from 1, 2…8. The time variables including DW, DM, and MY, respectively are the day of the week, day of the month, and month of the year. We include year and cities as fixed effects to build the panel model. This means that we estimate the effect of changes in meteorological and traffic conditions within cities on changes in air pollution.\par

The Root Mean Squared Error (RMSE) is commonly employed to quantify prediction error. In every scenario, smaller RMSE values signify reduced prediction error. We utilize the RMSE loss function for the evaluation and visualization of the losses.

\section{Results}

\subsection{Feature selection}
\subsubsection{Traditional statistics-led feature selection}

Figure 2 illustrates the monthly pollutant levels in three cities. Among the three cities, it reveals a consistent trend with NOX, wherein pollutant concentrations decrease from January to July, followed by an increase. Usually, the summer months (June to August) exhibit higher temperatures, leading to a decrease in pollutant concentrations, while winter months (December to February) see an increase. In contrast, $PM_{2.5}$ shows less seasonal variation and remains relatively evenly distributed. We report the correlation results from the panel model:  Pollution = f (Traffic, Weather)  presented in Figure 3. The full form of the variables' names can be referred to in Table 1.  

\begin{figure}[H]
    \centering
    \includegraphics[width=0.75\linewidth]{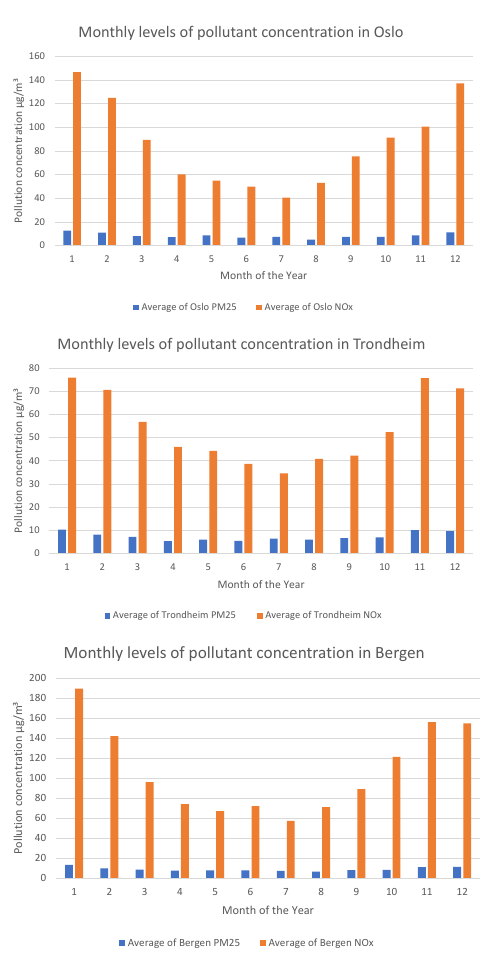}
    \caption{Monthly levels of two pollutants across the three cities. }
    \label{fig:monthly_levels}
\end{figure}

\begin{figure}[H]
    \centering
    \includegraphics[width=1.0\linewidth]{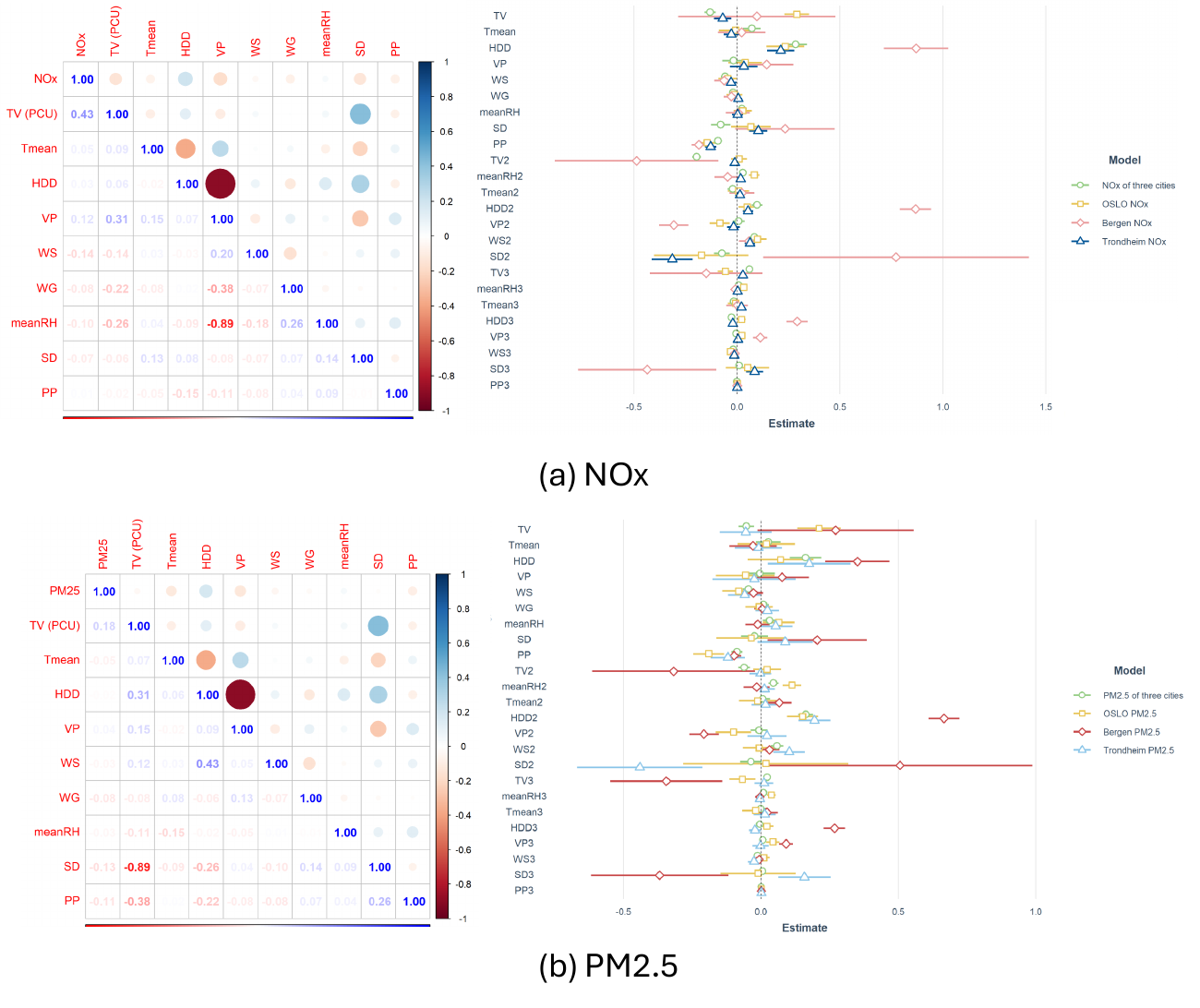} 
    \caption{The correlation results from the panel models at individual city and combined scale (a) for NOX and (b) for PM2.5. \texttt{Note: The full form of variables is presented in Table 1. The text in the image may appear small and may require zooming in for better visibility.}}
    \label{fig:galaxy}
\end{figure}

We initially exclude quadratic and cubic terms, focusing solely on examining the correlations between individual variables. According to the correlational matrices presented in Figure 3, we identified that VP (vapor pressure), meanRH2 (mean relative humidity over 24 hours), TV (traffic volume of passenger car units), and WS (average mean wind speed) exhibit the most statistically significant correlations with $NO_{x}$. Similarly, for $PM_{2.5}$, we observed statistically significant correlations with TV (traffic volume of passenger car units), SD (snow depth), and PP (precipitation). Therefore, TV emerges as a common variable for both pollutants, suggesting that increased traffic volume may lead to elevated air pollution concentrations.

Subsequently, we introduce our panel model for analysis. For both PM2.5 and NOX, we found that HDD and the square of SD demonstrate the highest correlations. Additionally, for PM2.5, PP and the square of HDD also exhibit significant correlations. Thus, the common variables for both pollutants include the square of HDD and the square of SD.\par

\subsubsection{ML-led feature selection}

We use three different ML approaches for feature selection to identify risk factors from traffic, weather, and air pollution. Feature selection is obtained by identifying important features through a selection of feature combinations rather than the selection of single features \cite{gromping2009variable}, \cite{kodinariya2013review}, \cite{murtagh2012algorithms}. Hence, we adopted different clustering methods for feature combination identification that contribute to increasing air pollution concentrations. The results are in Figures 4 to 6, and Appendix B. 

We began with two basic classes based on the data's structure and similarities. Using visualization, we categorized the data into two to four groups, revealing distinct patterns and separations among clusters or groups. We selected the number of two-, three-, and four-category clusters respectively. However, we found that Vapour Pressure (VP), Precipitation (PP), Wind Gust (WG), and mean air Temperature (Tmean) always clustered together, and the other variables clustered in another group with air pollution. These three weather variables have little correlation with the air pollutants studied in this paper. The two pollutants become one group with mean Relative Humidity (meanRH), Heating Degree Days (HDD), Snow Depth (SD), and Traffic Volume (TV),  average mean Wind Speed (WS), representing the four weather variables most correlated with pollutants. When divided into three or four clusters, HDD, SD, and TV are still clustered into one category, indicating that these four weather variables are correlated. Upon inclusion of quadratic and cubic terms, we discover that mean Relative Humidity (meanRH), mean Temperature (Tmean) cubed, Heating Degree Days (HDD), and the cube of HDD form a cluster with them. This means that those climate feature combinations have the greatest impact on air pollution. 


\begin{figure}[H]
    \centering
    \includegraphics[width=0.75\linewidth]{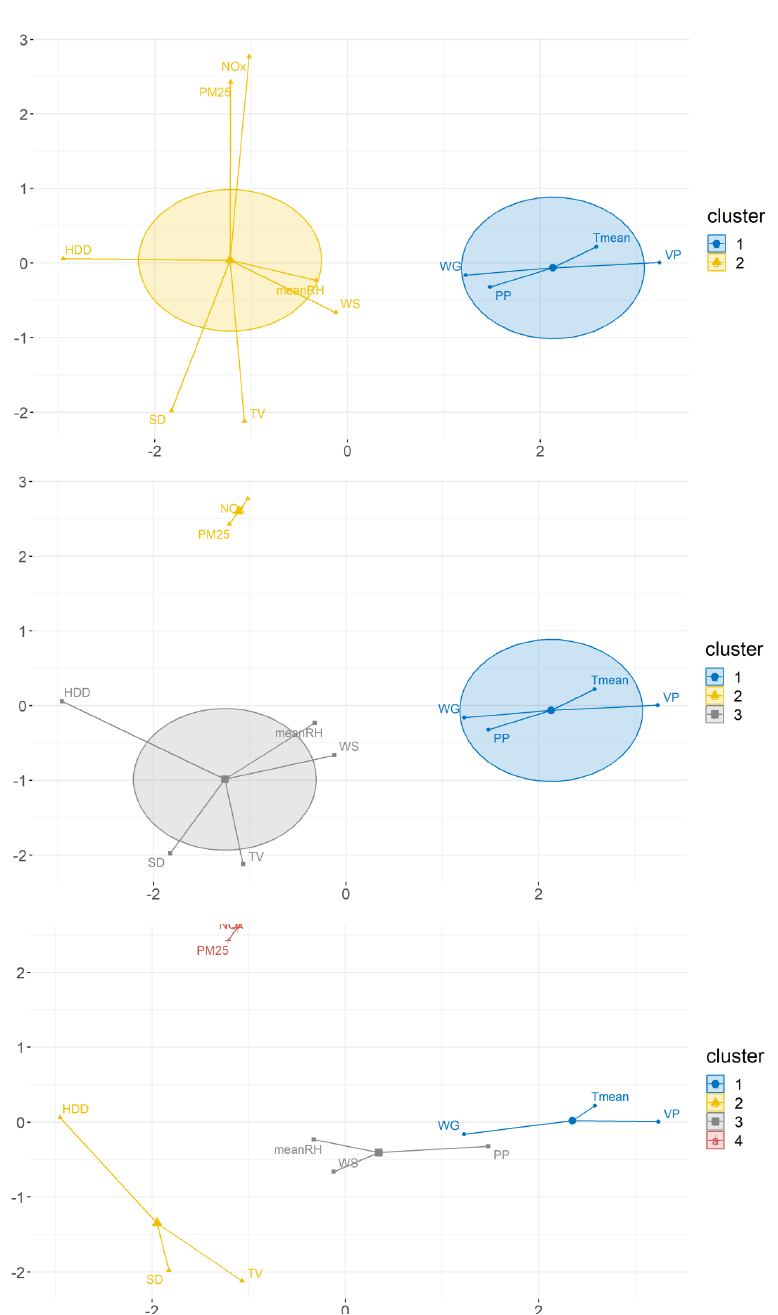}  
    \caption{Cluster plots generated using K-means. Note: The text in the image may appear small and may require zooming in for better visibility.}
    \label{fig:k_3_clusters}
\end{figure}

\begin{figure}[H]
    \centering
   \includegraphics[width=0.9\linewidth]{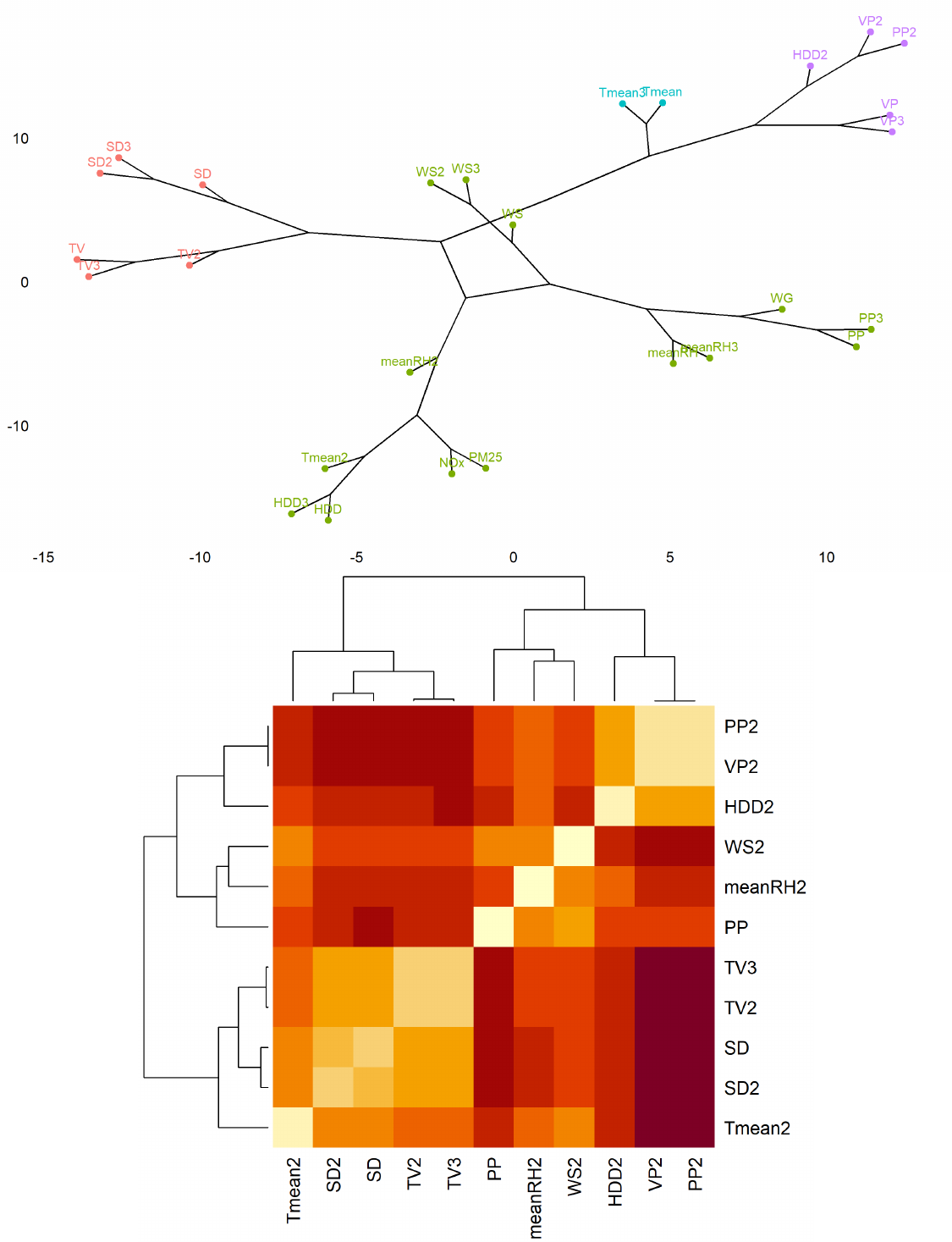}
    \caption{Hierarchical clustering plot after data normalization with no quadratic and cubic variables. The full form of the variables can be inferred from Table 1.}
    \label{fig:galaxy_1}
\end{figure}

Figure 6a and Figure 6b demonstrate the random forest results. To illustrate the results of the random forest analysis, the IncMSE percentage, representing the percentage of mean square error, serves as an indicator of variable importance. It signifies the extent to which the removal of a variable would decrease prediction accuracy. Consequently, a higher IncMSE percentage suggests greater importance of the variable. For example, $PM_{2.5}$ across the three cities in Figure 6, was consistently identified as the most important factor for influencing the HDD. Similarly for $NO_{x}$, HDD and PP emerged as the most influential variables. Thus, according to the random forest analysis results, the variable that played a decisive role in the prediction performance of air pollution concentration ($NO_{x}$ and $PM_{2.5}$) is HDD.

\begin{figure}[H]
\centering
\subfloat[$NO_{X}$]{%
  \includegraphics[width=1.0\linewidth]{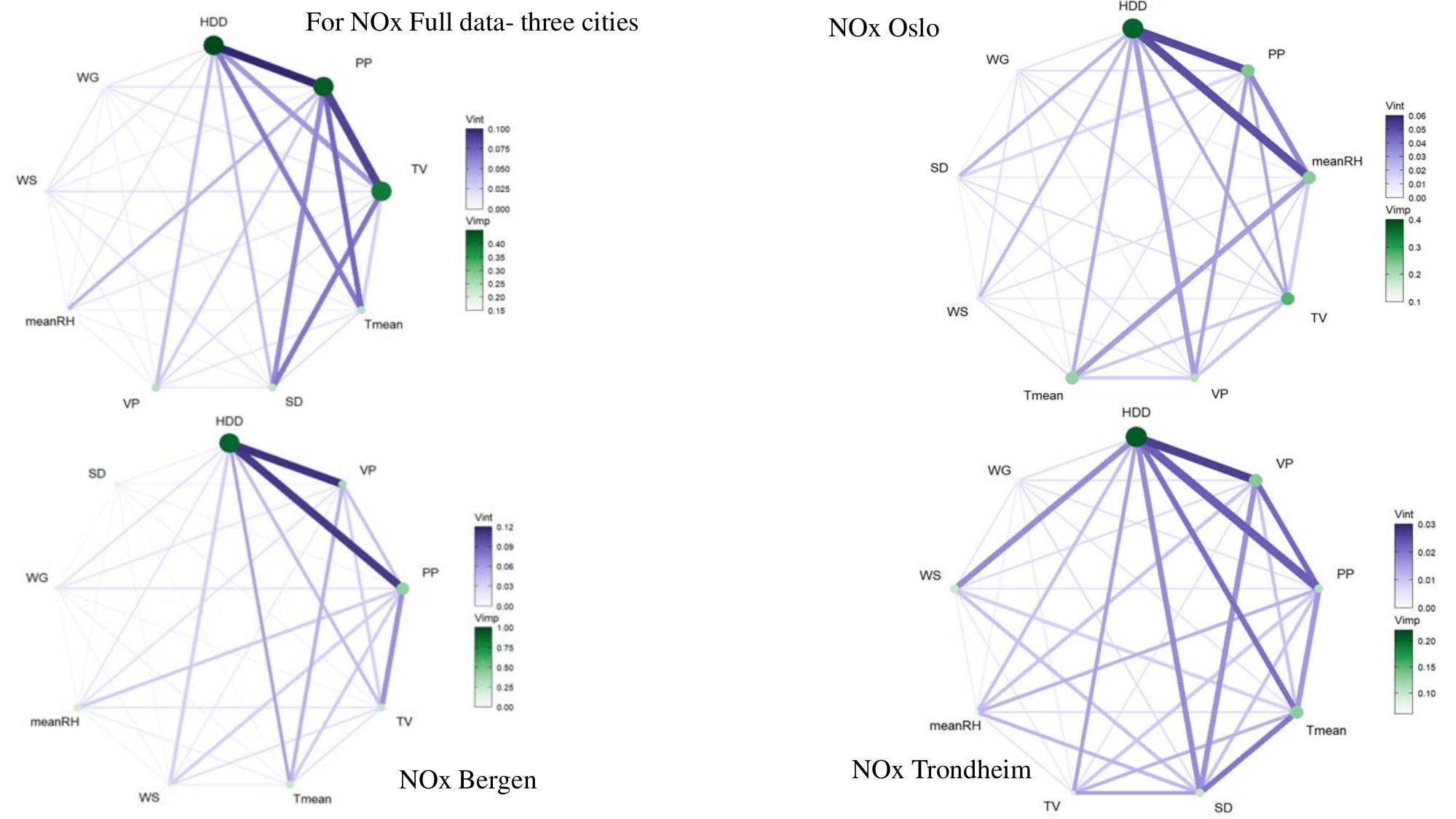}%
}

\subfloat[ $PM_{2.5}$]{%
  \includegraphics[width=1.0\linewidth]{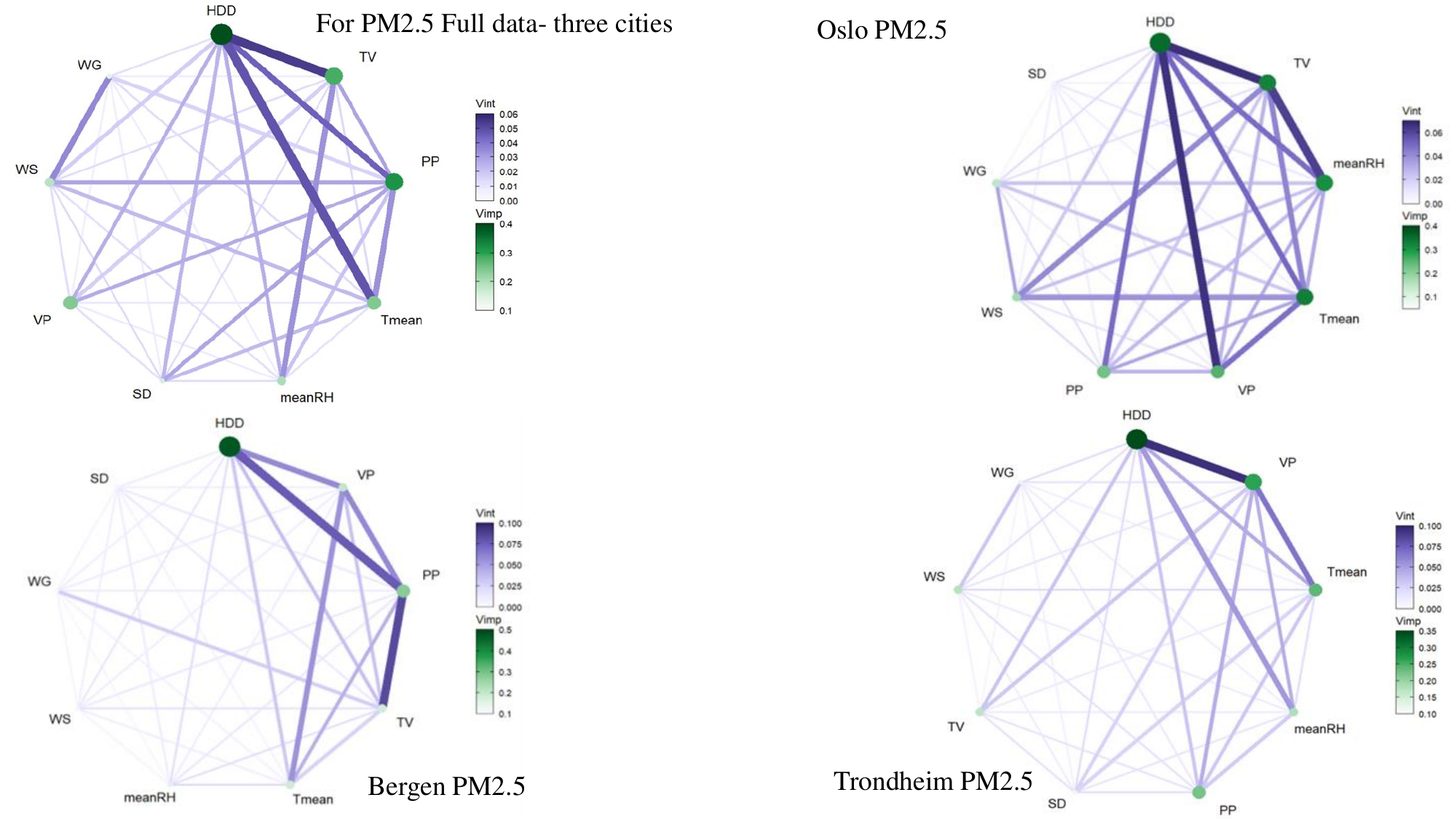}%
}

\caption{Random forest results illustrating with Variable Importance (Vimp) and Variable Interaction Effects (Vint) for (a) $NO_{X}$ and (b) $PM_{2.5}$. [Note: \texttt{The variables include Traffic Volume (TV), mean Temperature (Tmean), Heating Degree Day (HDD), Vapor Pressure (VP), mean Wind Speed (WS), mean Wind Gust (WG), mean Relative Humidity (meanRH), Snow Depth (SD), and Precipitation (PP).}]}
    \label{fig:galaxy_2}
    
\end{figure}

In traditional approaches, Traffic Volume (TV), Snow Depth (SD), and Heating Degree Days (HDD) are significant factors affecting both pollutants. Meanwhile,  the K means and hierarchical clustering’s feature selection results show meanRH and HDD have a direct impact on air pollution. Snow depth is also associated with air pollution concentration. The random forest results show that the variables that play a decisive role in the prediction performance of air pollution ($NO_{x}$ and  $PM_{2.5}$) are HDD. The variable mentioned in both machine learning and traditional statistical approaches is HDD, which emphasizes its key role in the impact on air pollution, and an increase in the number of heating degree days is associated with air pollution, which represents an increase in days with temperatures below 17 degrees. 

From the monthly variation of HDD and $PM_{2.5}$  in this decade for three Norwegian cities in Figures 15 and 17 Appendix C. We find that HDD gradually increases from October onwards, followed by a decrease from January onwards. The changes in HDD are mainly due to local air temperature changes; the concentration of $PM_{2.5}$ increases significantly from October and begins to decrease from January, with similar changes to HDD. Therefore, the number of HDD in winter relates to air pollution. \par 
In November, $PM_{2.5}$  concentrations in Trondheim are higher than the average of the three cities, with the largest change from October to November. Figure 16 shows the variations in HDD and  $PM_{2.5}$ in Trondheim, and we find that they have a consistent trend. When HDD decreases from January to May,  $PM_{2.5}$ concentration follows suit, and conversely, when HDD increases from October to December,  $PM_{2.5}$ concentration also rises. Figures 12 to 14 in Appendix C, demonstrate congruent findings. Taken together, this suggests that increased heating activity may contribute to the increase of $PM_{2.5}$ levels.\par 

\subsection{Predictions based on deep learning approaches}

To evaluate the prediction performance of deep learning, the deep learning algorithms we use here are LSTM and PBDL. We use both algorithms in the deep learning library Keras\footnote [5]{Code environment: python 3.8.2, TensorFlow 2.1.0, Keras 2.3.1} for multivariate time series forecasting, we exclude the time variable and select traffic and weather variables to predict $PM_{2.5}$ and $NO_{x}$ respectively. We have a total of ten variables. This algorithm uses the traffic and eight weather variables and the $PM_{2.5}$ or $NO_{x}$  concentration from the previous days, to predict $PM_{2.5}$ or $NO_{x}$  concentration on the next day. We use the two deep learning algorithms to run 3 cities' data independently. Both models underwent hyperparameter tuning using the most optimal parameters. The results shown are in Table 2. Additionally, the process of optimizing the algorithm, hyperparameter tuning, and presenting detailed results can be found in Appendix D Tables 6 to 8. Figures 18 to 20 in Appendix D visually depict the loss reduction curves of the two algorithms across the three cities. Appendix D figures 21 to 23 show the actual and predicted plots.\par

 \begin{table}[H]%
 \centering
 \begin{threeparttable}
	\caption{Performance comparison between LSTM and PBDL}
  \label{Performance comparison between LSTM and PBDL}
	\centering
	\begin{tabular}{ccc}
      \hline
\textrm{Cities}&
\multicolumn{1}{c}{\textrm{LSTM'  RMSE}}&
\textrm{PBDL' RMSE}\\
		\hline
Oslo $NO_{x}$	&0.7640 &0.7458\\
Bergen $NO_{x}$ 	&1.0190	&1.0442 \\
Trondheim $NO_{x}$ 	&0.4430	&0.4424 \\ 
Oslo $PM_{2.5}$ 	&1.0970	&1.0866 \\
Bergen $PM_{2.5}$	&0.7460	&0.7201 \\
Trondheim $PM_{2.5}$   	&1.0320	&0.4448 \\
  \hline
    \end{tabular}
    \end{threeparttable}
\end{table}
\par

From the results, when compared with LSTM, except for Trondheim’s $NO_{x}$ and Bergen’s $NO_{x}$, PBDL exhibits superior prediction performance. The daily average for $PM_{2.5}$ is 10.88 $\mu g/m^3$, with a maximum value of 53. Taking Trondheim as an example, when the accuracy increases from 1.0320 to 0.4448, it means that PBDL can more accurately predict an average concentration of 6.39 $\mu g/m^3$ ($10.88 \times (1.0320 - 0.4448)$), or a maximum of 31.12 $\mu g/m^3$ ($53 \times (1.0320 - 0.4448)$) of growth. 

This increase in predictive performance has strong policy implications for air pollution governance. For example, in Norway, before 2016, there existed only a limit value for the annual average $PM_{2.5}$ concentration, which must not surpass 15 µg/m³. The Pollution Control Regulations tightened the limit values for particulate matter in January 2016. Presently, the daily limit for $PM_{2.5}$ is 10 micrograms per cubic meter, with an annual limit value of 25 µg/m³, in line with air quality standards set by the European Union \cite{ortis2020potential}. 

In the case of $NO_{x}$ predictions, Trondheim shows the smallest predicted RMSE, Oslo with an intermediate, and Bergen with the highest values. Conversely, for $PM_{2.5}$ results, Bergen showed the smallest predicted RMSE, while Oslo had the highest. While LSTM outperforms PBDL in predicting Bergen's NOx levels, even marginal improvements guide attention toward tailored climate policy adjustments. Bergen is characterized by persistent heavy rainfall and frequent storms throughout the year. Appendix A Tables 4 and 5 illustrate the climate and energy policies and accomplishments in Bergen during our study period. We found that during the period from 2009 to 2014, Bergen implemented air quality policies, leading to a drop in air pollutants after 2010.$NO_{x}$ and $PM_{2.5}$ levels in Bergen remained relatively stable after 2010, while concentrations of these pollutants in Oslo and Trondheim continued to decline after 2015. However, $NO_{x}$ levels increased in Trondheim in 2018. Consequently, despite improvements in air quality in the three cities during policy implementation, no correlation was observed between policy implementation and the predictive performance of deep learning algorithms, which highlights the complex interplay between policy measures and air quality improvements. \par

Given that a single deep learning algorithm exhibits varying prediction accuracies across different cities, and these algorithms undergo optimal parameter selection and employ diverse methods to mitigate overfitting, while the datasets from these cities possess comparable data quality, feature distributions, sample sizes, and so forth, the plausible explanation lies in the differing distribution and variation of data values, thereby leading to distinct data complexities. \cite{bailly2022effects}. \par

\section{Discussion}

In this study, we identified the primary climatic factors that influence air pollution and examined the predictive advantages and policymaking utility presented by the Long-term Short-Term Memory (LSTM) deep learning model and the Physics-Based Deep Learning (PBDL) approach. In doing so, our approach involves the utilization of a data-driven and multi-method methodology, leveraging 10 years of daily data spanning three major cities in Norway. We incorporate traffic data, ten weather variables, time-related variables, and data on two key air pollutants for comprehensive analysis.\par

By comparing the feature selection results between the panel model and machine learning algorithms, specifically, K-Means (KM), Hierarchical Clustering (HC), and Random Forest (RF), we observe a strong correlation between heating degree days (HDD) and the concentration of both pollutants. In addition, we find that HDD shares a consistent seasonal trend with air pollutant concentrations. This suggests that winter heating in Norway contributes to the exacerbation of air pollution. We find that the concentration of $PM_{2.5}$  increased in winter. This also reflects a certain extent that air pollution is related to the widespread use of wood for heating in Norway in winter. \par


Notably, PBDL exhibits the highest accuracy in predicting PM2.5 levels in Trondheim, achieving an increased accuracy of 6.39 $\mu g/m^3$. Despite Norway's environmentally conscious reputation, current PM2.5 levels in these cities exceed both European Union and Norwegian standards \cite{ortis2020potential},  emphasizing the urgency of accurate forecasting for public health protection \cite{madsen2020can}, \cite{council200850}. Controllable variables such as Traffic Volume (TV) and Heating Degree Days (HDD) highlight policy intervention opportunities.\par

Moreover, according to our prediction results, PBDL demonstrates superior forecasting capabilities compared to LSTM for the average concentration of $PM_{2.5}$ in Trondheim, achieving an accuracy of 6.39 $\mu g/m^3$. The current daily average of $PM_{2.5}$ in these three Norwegian cities stands at 10.88 $\mu g/m^3$, already exceeding both the European Union and Norway's PM2.5 limit standards \cite{ortis2020potential}. These standards indicate potential harm to public health if exceeded 
 \cite{madsen2020can}. Hence, a marginally more precise forecast holds significant importance as it directly impacts human health, posing potential harm, especially for Trondheim.\par

Since LSTM performs better than PBDL in predicting Bergen's NOX levels, even with only a slight improvement, it guides our attention towards exploring climate policy changes particular to Bergen. Based on the findings in Appendix A Tables 3 to 5, it was observed that a new air quality action plan was formulated in 2010. Despite this, Oslo continues to surpass the Air Quality Directive 2008/50/EC, 2008 (AQD) limits for both annual mean and hourly $\text{NO}_2$
 concentrations \cite{council200850}. Bergen's predictions for the two pollutants differ significantly, directing our attention towards examining climate policy innovations specific to Bergen. This involves integrating the outcomes of deep learning models with the implementation of policy measures. Bergen is characterized by persistent heavy rainfall and frequent storms throughout the year, as documented by Bergen Kommune (2018: 8–9). Appendix A Tables 4 and 5 illustrate the climate and energy policies and accomplishments in Bergen during our study period \footnote{Green strategy climate and energy action plan for Bergen. (n.d.). 'Gasoline Prices around the World, 28-Sep-2015 | GlobalPetrolPrices.com'$http://www.globalpetrolprices.com/gasoline_prices/$ [accessed 5 October 2015]}. We found that during the period from 2009 to 2014, Bergen implemented air quality policies, leading to a drop in air pollutants after 2010.$NO_{x}$ and $PM_{2.5}$ levels in Bergen remained relatively stable after 2010, while concentrations of these pollutants in Oslo and Trondheim continued to decline after 2015. However, $NO_{x}$ levels increased in Trondheim in 2018. Consequently, despite improvements in air quality in the three cities during policy implementation, no correlation was observed between policy implementation and the predictive performance of deep learning algorithms. \par

\section{Conclusion}

Conventional wisdom suggests that LSTM should be  the go-to deep learning algorithm for handling time series data\cite{karim2017lstm}. Yet, in this study, PBDL has not only challenged but surpassed LSTM in this specific case. In our study, we use weather and air pollution variables such as traffic flow, average temperatures, and heating degree days. These variables are associated with chemical and dynamical processes in the atmosphere. For example, traffic flow can affect the spread of pollutants; average temperature and humidity can affect the rates of chemical reactions in the atmosphere; and the wind speed and wind gusts can control the transmission and spread of pollutants. Therefore, we can see the advantages of PBDL that arise from capturing physical information in the model. In contrast, traditional deep learning models such as LSTM usually make predictions by simply learning patterns in time series data, lacking an understanding of the physical mechanisms behind the data. Therefore, PBDL is more likely to benefit when the data has clear physical regularities, while LSTM may perform better when the data does not. In summary, PBDL performs better when processing time series data because it is better able to combine data and physical equations, thereby more accurately capturing the dynamic behavior of the system. For data without clear physical laws, LSTM may be more suitable because it is better at learning patterns and long-term dependencies in the data.\par

Furthermore, our analysis uncovers the intricate interactions between pollution and meteorology in feature selection (Mirzavand et al., 2023). By predicting various air pollutants, we significantly enhance the accuracy of single pollutant forecasts by simulating the complex physical and chemical processes involved, while incorporating essential meteorological factors such as wind speed,  wind gusts, air pressure, heating degree days, traffic volume, and snow depth. The inclusion of these essential meteorological factors, proven to be crucial in air quality dynamics, refines and optimizes our model, ensuring the most effective simulation of real-world phenomena (Xu et al., 2022).\par

Our meticulous integration of the timeline of Norwegian air quality policies and the occurrence of the Bergen climate disaster event aimed to explore the impact of environmental policies on the accuracy of air pollution predictions (Zhao et al., 2023). Despite not identifying significant impacts of policy changes on prediction accuracy, our pioneering research combines physics-based deep learning with fluctuations in air quality policies.\par

Our study is limited by the absence of detailed spatial location information and our study largely focuses on the temporal dimension in our data. Consequently, we incorporated Ordinary Differential Equations (ODE) into the PBDL framework instead of Partial Differential Equations (PDE) to address this limitation. To overcome this constraint, future research efforts could focus on gathering datasets that include both temporal and spatial dimensions. This would allow for a more comprehensive utilization of PBDL, thereby enhancing its ability to analyze and understand complex systems more effectively.\par

While LSTM models have found extensive use in environmental research for forecasting and scenario analysis, the utilization of PBDL in policy analysis is an emerging research field, requiring further exploration to understand its potential advantages and limitations. In conclusion, our findings underscore the critical role of climate factors and deep learning models in air pollution prediction. Reliable and exceptionally precise prediction models play a pivotal role in guiding urban policies. Tailored policy interventions, guided by accurate forecasting, are essential for mitigating the adverse impacts of air pollution on public health and environmental sustainability. \par


\section*{Author Contribution Statement}\par
CC, RD, and RMA conceptualized the research. CC performed data collection, data pre-processing, methodological development, formal data analysis, and interpretation of the results. CC, RD, and RMA wrote the manuscript. RD and RMA reviewed and edited it. RMA provided the overall supervision and the funding acquisition. All authors have read and agreed to the published version of the manuscript. 

\section*{Funding}\par
CC and RMA acknowledge funding support from the Caltech Center for Science, Society, and Public Policy. RD acknowledges funding support from Cambridge Arts, Humanities and Social Science (AHSS) Research Grants 2023-24, Keynes Fund [JHVH], and Bill and Melinda Gates Foundation [OPP1144].\par

\section*{Data and code availability statement}
The traffic flow, air pollutants, and meteorological data used in this work are publicly available and freely available from the Norwegian Public Road Administration, Statens Vegvesen (SVV), Norwegian Institute for Air Research (NILU), and the Norwegian Meteorological Institute (MET). The descriptive statistics for this data can be located in section 2.\par
The code is made available from 
\href{https://github.com/congca/Physics-based-DL-reveals-rising-heating-demand-heightens-air-pollution-in-Norwegian-cities/tree/main}{Github repository}.
To acquire the integrated datasets, please send a request to congc@caltech.edu.\par
\section*{Acknowledgments}
We thank Zongyi Li from Caltech for providing us with invaluable advice.

\section*{Conflicts of Interest}\par

The authors declare no conflict of interest.\par

\bibliographystyle{apacite}
\bibliography{ref}
\section*{Appendix A Norwegian Air Quality Policies and Weather Events }\par
 
Norwegian authorities have been proactive in addressing air quality issues, with a series of influential air quality action plans over the past two decades.  For example, the Norwegian transport-related air quality policy focuses on reducing air pollution caused by transportation, as in Norway, traffic is a significant contributor to pollution, with PM and $NO_{x}$ identified as the most important pollutants in Norwegian cities. Norway implemented a series of air quality improvement policies in the 14 years from 2004 to 2019. As our research spans from 2009 to 2018, we conducted a review of these policies, considering the legal and public opinion context in which the Norwegian policies were implemented. \par

\begin{table}[H]
    \centering
    \begin{threeparttable}
\footnotesize

\caption{Air Quality Policies in Norway}
\label{tab:air_quality_policies}
\begin{tabular}{p{2.7cm}p{1.2cm}p{0.55cm}p{9.5cm}}
\hline
Authorities & Coverage area & Year & Policies \\
\hline
Norwegian Institute for Air Research (NILU)&Norway& 2004 &Norway follows the existing policies outlined in the Clean Air for Europe (CAFE) Directive (EP and CEU, 2008) and (EP and CEU, 2004).\\
NPRA East Region branch and NILU&Oslo & 2004 & First air quality action plan, improve the $NO_{2}$ concentrations \\
& Oslo & 2011 & Reduce annual air quality action plan  \\
& Norway & 2015 & Following a ruling by the ESA (European Free Trade Association - EFTA - Surveillance Authority), efforts are underway to bring average  $NO_{2}$  concentrations below the Air Quality Directive (AQD) annual average limit \cite{council200850}, involving measures such as reducing vehicle speed, incentivizing electric vehicles, and decreasing traffic flow \cite{amundsen2018low}. However, the main roads in Oslo persist in exceeding the standard. \\

&  &  & The government discouraged private car usage and implemented a national transportation plan from 2014 to 2023 to promote public transportation initiatives. \\
&  &  & Counties received significant yearly funding of 12 billion Norwegian kroner, along with various incentives  \\
&  &  &  Exemptions from value-added tax and purchase tax for electric vehicles were implemented to encourage non-motorized transportation\footnote[6]{Norway Air Quality Policies (2015). ‘Gasoline Prices around the World, 28-Sep-2015 | GlobalPetrolPrices.com’ <$http://www.globalpetrolprices.com/gasoline_prices/$> }. \\
& Norway & 2016 & The law permits the imposition of congestion charges and regional charges based on emission classes (§ 27, Lov om vegar) \\
&  &  & Norway decided to lower its particulate matter limits, making Norway's policy more stringent than that of the EU\\
Norwegian Ministry of Transport & Oslo & 2017 & A new toll policy was introduced, where road tolls varied based on vehicle fuel type, as well as peak and off-peak hours  \\

& Norway & 2017 & Zero-emission vehicles were permitted to pass through toll plazas free of charge. The action plan mentioned the need to reduce traffic volumes compared to 2015 to achieve emission reductions, setting a politically targeted reduction goal of 20 percent, but did not propose specific measures for traffic reduction  \cite{amundsen2018low} \\
& Bergen & 2017 & Environmentally friendly road tolls. Related policy also includes the speed limit policy, however, a study explored the impact of Norway's speed limit policy on local air pollution, and they concluded that reducing vehicle speeds did not reduce air pollution  \cite{WAERSTED2022100160} \\
& Oslo & 2018 & The reduction of on-street parking spaces, one-way streets, and the prohibition of private cars on some streets  \\
& Oslo & 2019 & Zero-emission vehicles will also need to pay tolls \\
&  &  & Fifty-three new road toll booths were opened in Oslo, with prices slightly lower than those in Oslo. Auto PASS drivers receive a 10 percent discount in Oslo and a 20 percent discount in Bergen. A 5 percent reduction in traffic volume compared with the same period last year  \cite{Oslo2015-2020} \\
\hline
\end{tabular}
   \end{threeparttable}
\end{table}
\par
 \begin{table}[H]
    \centering
    \begin{threeparttable}
\caption{Environmental Policies and Accomplishments in Bergen}
\label{tab:environmental_policies}
\begin{tabular}{c p{3.8cm} p{4.55cm}}
\hline
\textbf{Year} & \textbf{Policies} & \textbf{Accomplishment} \\ \hline
2009--2014 & The City of the Future program, which aimed to achieve minimum greenhouse gas emissions and adapt to climate change &  \\  
2013 & & Won the Urban Environment Award for developing a suburban light rail system that promotes sustainable development; Public transport usage saw an increase \\  
2015 & Completion of an underground waste collection system & Contributed to an enhancement in air quality; Bergen's adapted improved air pollution calculation model reduced emissions from traffic on busy roads.\\  \hline
\end{tabular} 
    \end{threeparttable}
\end{table}
 \par

\begin{table}[H]
    \centering
    \begin{threeparttable}
    \begin{tabular}{p{1.3cm}p{1.8cm}p{2.55cm}p{7.9 cm}}
        \hline
     
        \textbf{Area} & \textbf{Year} & \textbf{Weather Phenomenon} & \textbf{Climate Disaster} \\ \hline
        Bergen & January 2010 & Sustained temperature reversal & Worst air pollution incident: The hourly average NO$_2$ concentrations exceed 400 µg/m³ and surpass national air quality goals \cite{Wolf-Grosse2017}. \\
         & November and December 2010,January 2013 & Sustained temperature reversal & Hourly average concentrations of 150 µg/m³ were exceeded at least once per day over 11 or 12 days \cite{Wolf-Grosse2017}. \\ 
        Western Norway & October 2014 & Extreme rainfall & Floods and landslides occurred; however, most respondents believed the city wouldn't experience such extreme weather again, so they didn't see the need for significant preventive measures. \\ 
        Norway & December 2020 & Extreme rainfall & Severe landslides \cite{Amundsen2021}. \\ 
        & August 2023 & Extreme rainfall & Storm \cite{Amundsen2021}. \\ \hline
    \end{tabular}
    \caption{Summary of weather phenomena and climate disasters}
    \label{tab:weather-disasters}
   \end{threeparttable}
\end{table}

\section*{Appendix B Heatmaps and More ML-led Feature Selection Results }\par
Figures 7 employ heatmaps, whereas Figures 8 through 10 employ KM to explore the effects of quadratic and cubic variables on pollutants. Figure 11 utilizes HC, demonstrating identical outcomes.\par

\begin{figure}[H]
    \centering
   \includegraphics[width=0.8\textwidth, angle=270]{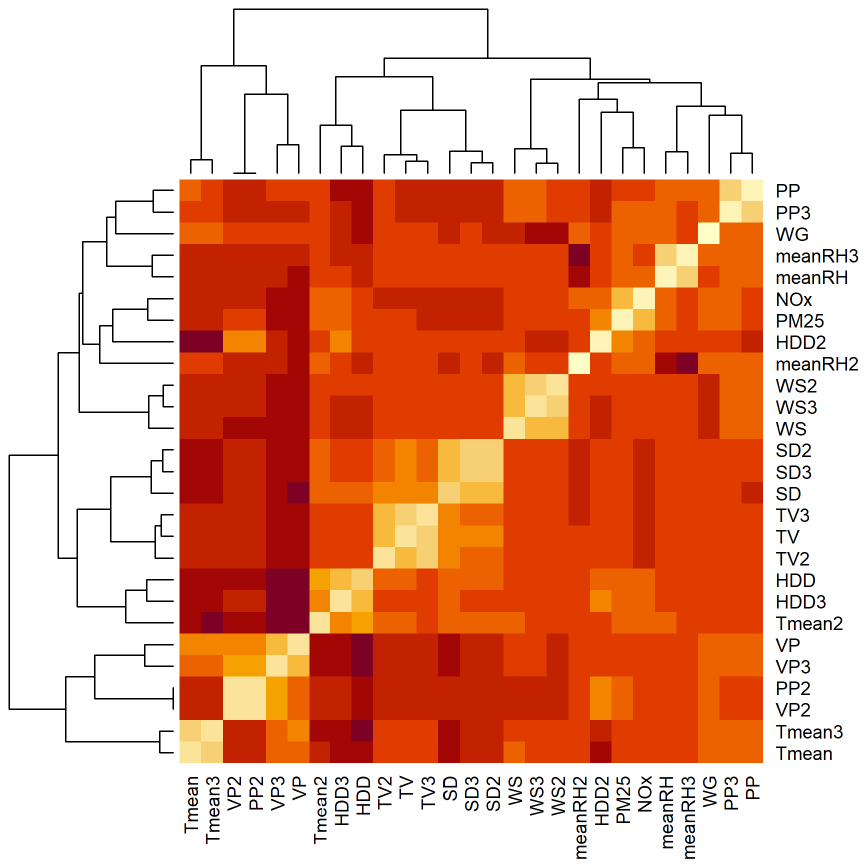}
    \caption{Heatmap.When there are quadratic and cubic variables, hierarchical clustering plots. Note: The text in the image may appear small and may require zooming in for better visibility.}
    \label{fig:galaxy_1}
\end{figure}

\begin{figure}[H]
    \centering
    \includegraphics[width=0.7\textwidth, angle = 270]{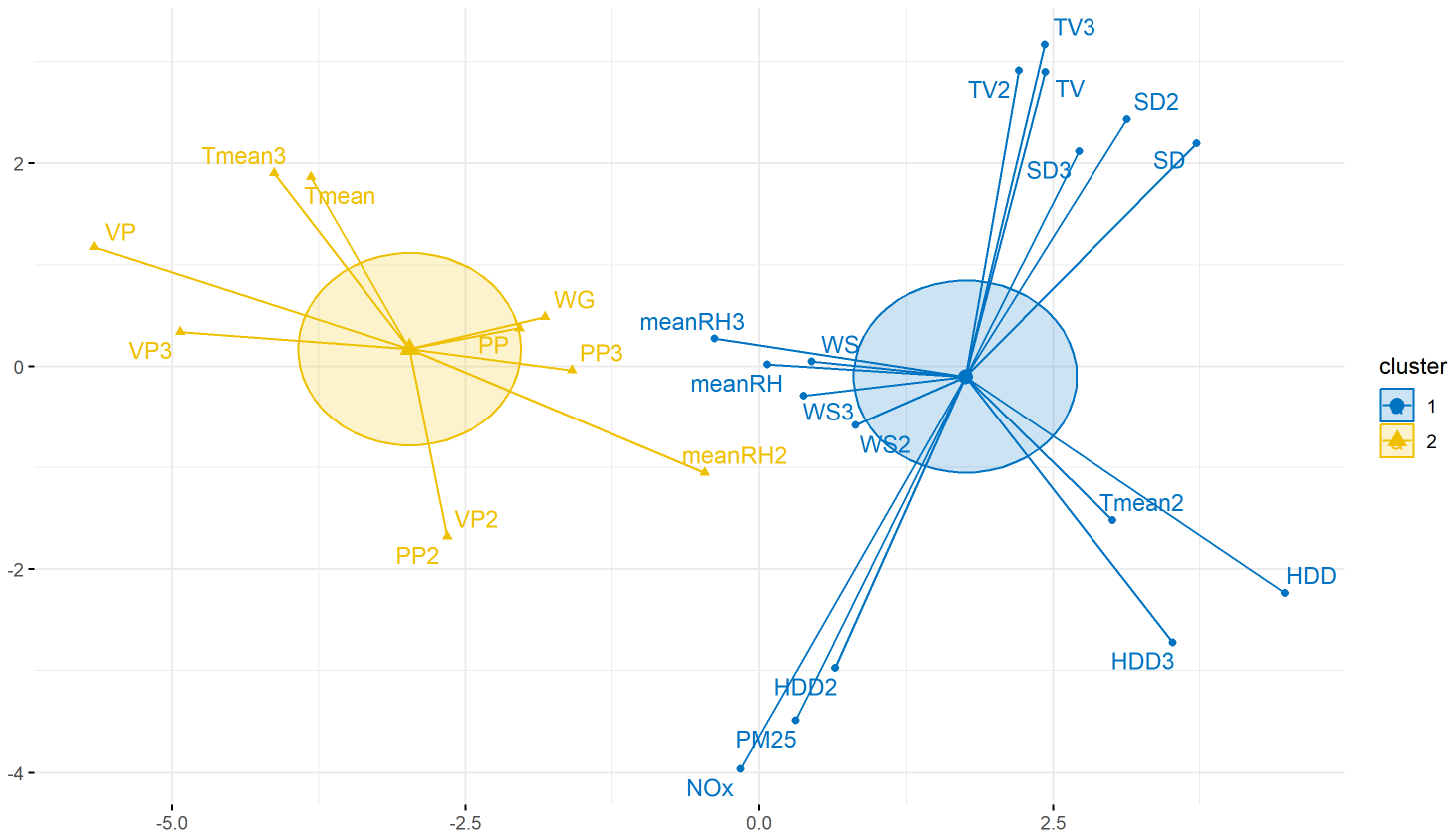}
    \caption{Cluster plots generated using K-means with two clusters}
    \label{fig:k_2_clusters}
\end{figure}

\begin{figure}[H]
    \centering
    \includegraphics[width=0.7\textwidth, angle = 270]{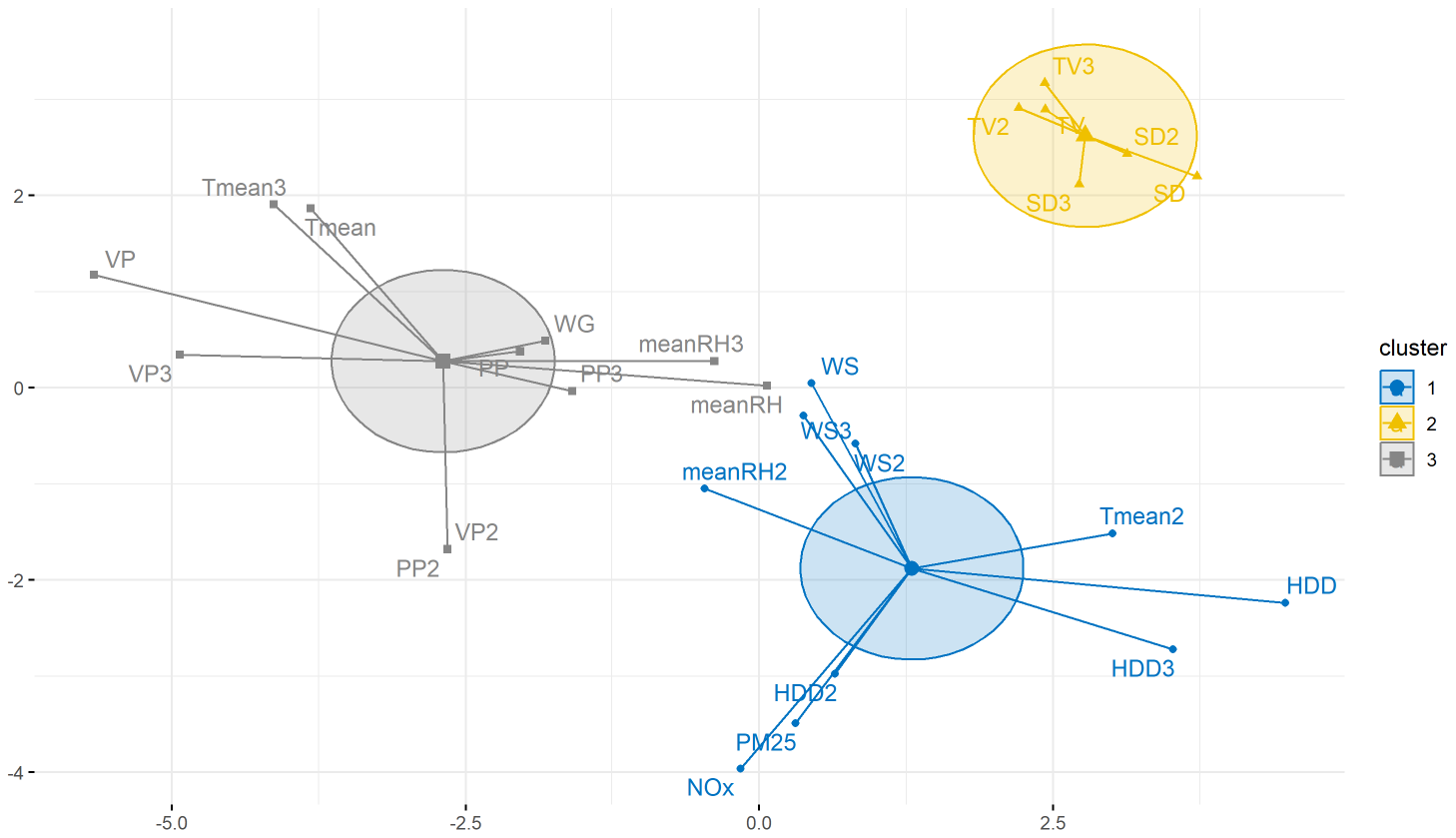}
    \caption{Cluster plots generated using K-means with three clusters}
    \label{fig:k_3_clusters}
\end{figure}

\begin{figure}[H]
    \centering
    \includegraphics[width=0.7\textwidth, angle = 270]{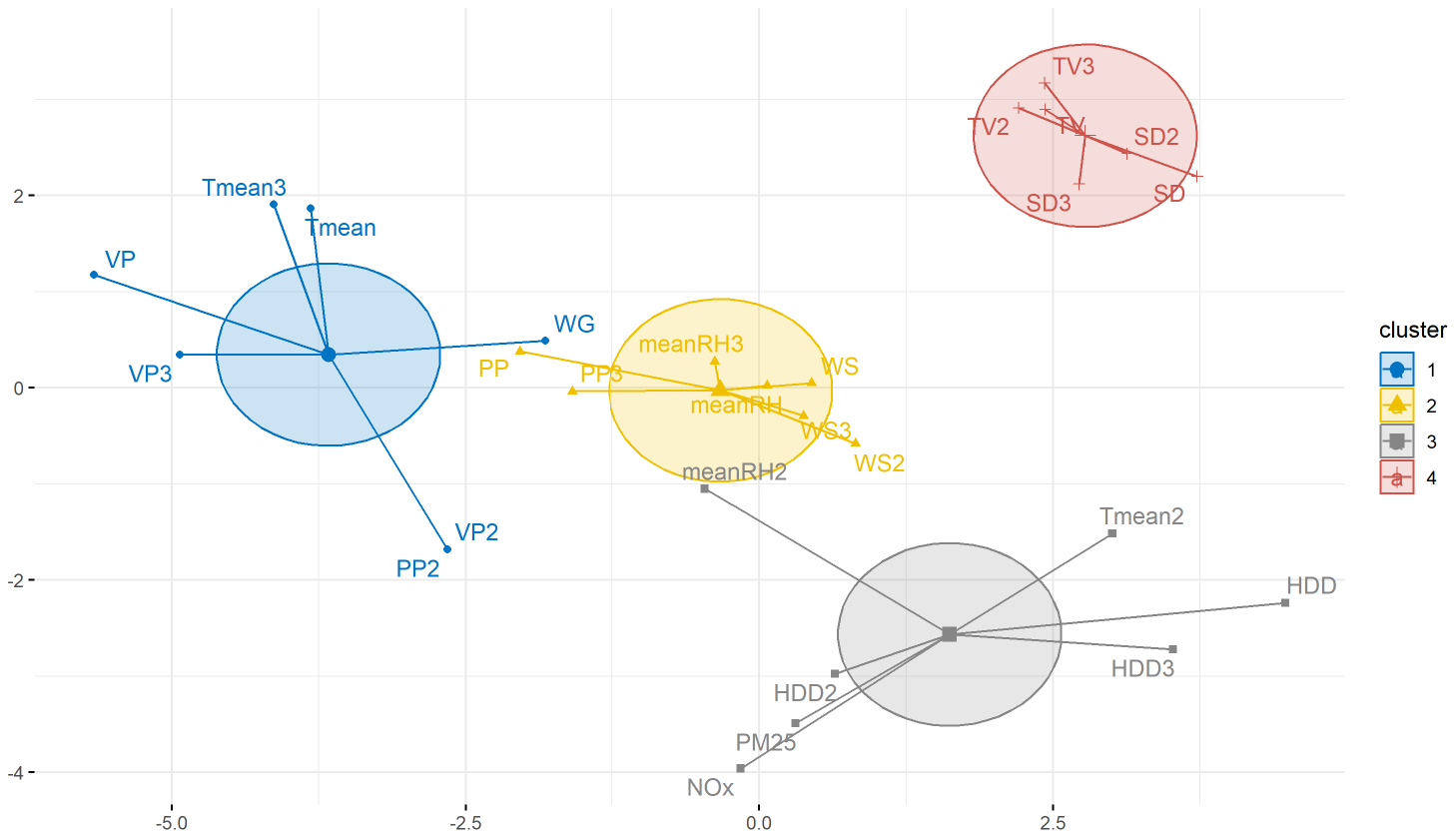}
    \caption{Cluster plots generated using K-means with two clusters}
    \label{fig:k_2_clusters}
\end{figure}

  \begin{figure}[H]
    \centering
   \includegraphics[width=\textwidth]{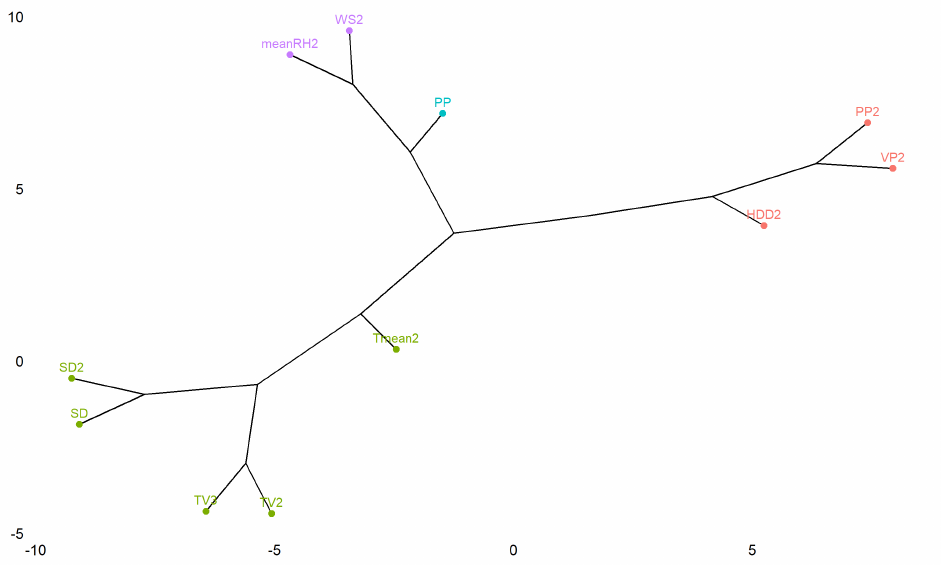}
    \caption{Hierarchical clustering plot after data normalization. Note: The text in the image may appear small and may require zooming in for better visibility.}
    \label{fig:galaxy_1}
\end{figure}

\section*{Appendix C Time Series Curves}\par
Figures 8 show the daily levels of two pollutants in three cities, covering ten years and totaling 3650 days. Notably, the $NO_{x}$ pollutant levels in Bergen consistently exceed those in Oslo and Trondheim, with many extreme values above 750. In contrast, there is minimal variation in $PM_{2.5}$ levels across the three cities. 

Combining the annual levels of the two pollutants in Figure 10, we found that the $NO_{x}$ concentration in Bergen was higher than 150 in 2010, while it decreased significantly in subsequent years. Trondheim exhibited the lowest $NO_{x}$ concentration.

\par

\begin{figure}[H]
    \centering
    \includegraphics[width=\linewidth]{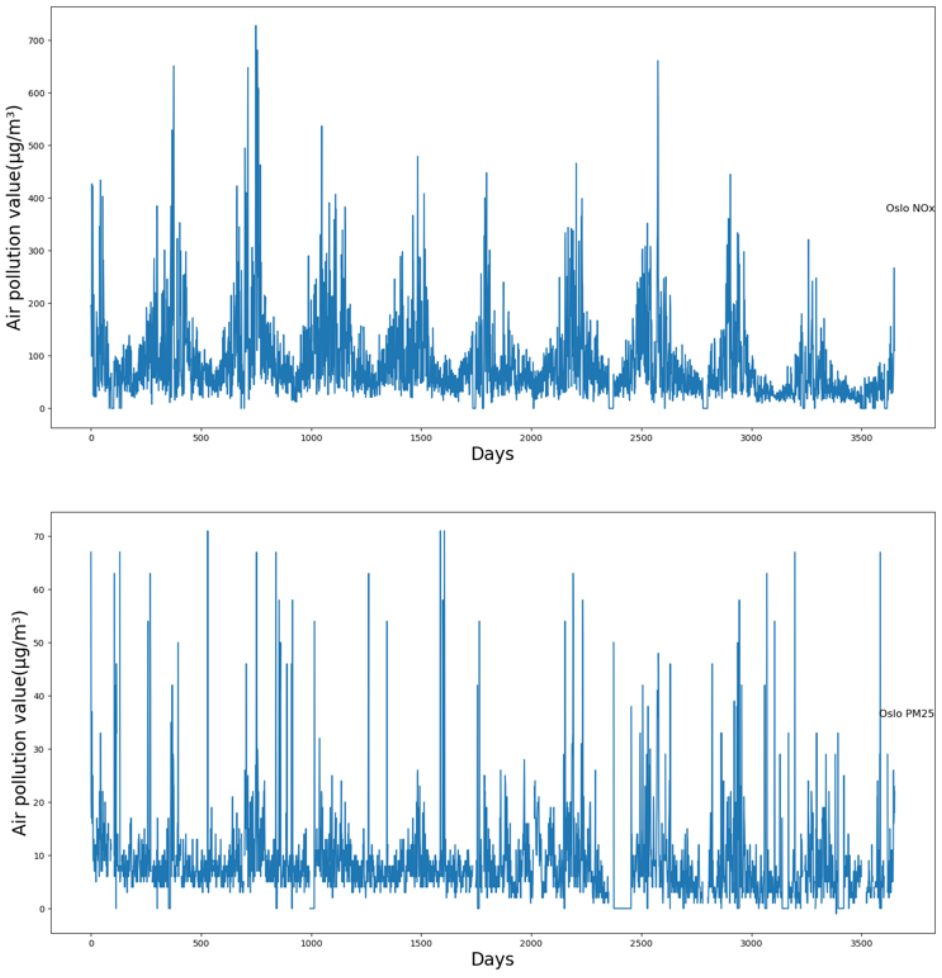}
    \caption{Daily levels of two pollutants: Oslo}
    \label{Daily}
\end{figure}

\par

\begin{figure}[H]
    \centering
    \includegraphics[width=\linewidth]{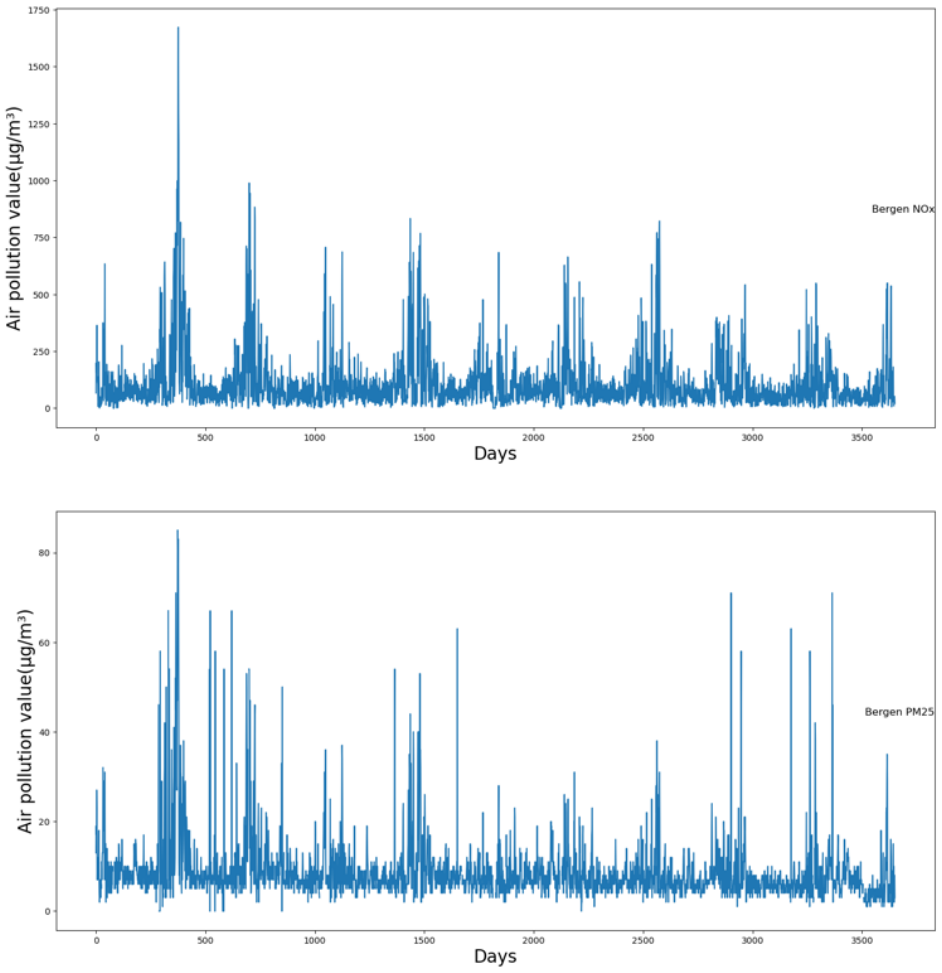}
    \caption{Daily levels of two pollutants: Bergen}
    \label{Daily}
\end{figure}

\par

\begin{figure}[H]
    \centering
    \includegraphics[width=\linewidth]{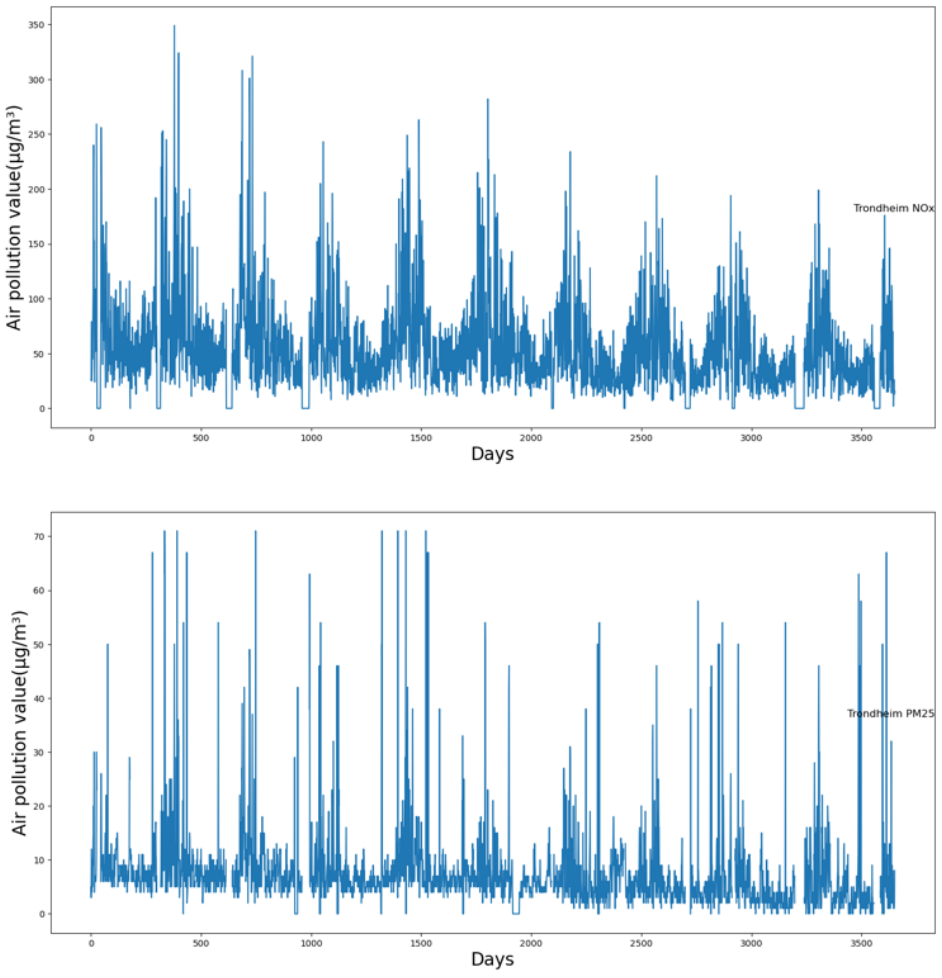}
    \caption{Daily levels of two pollutants: Trondheim}
    \label{Daily}
\end{figure}

\begin{figure}[H]
    \centering
    \includegraphics[width=\linewidth]{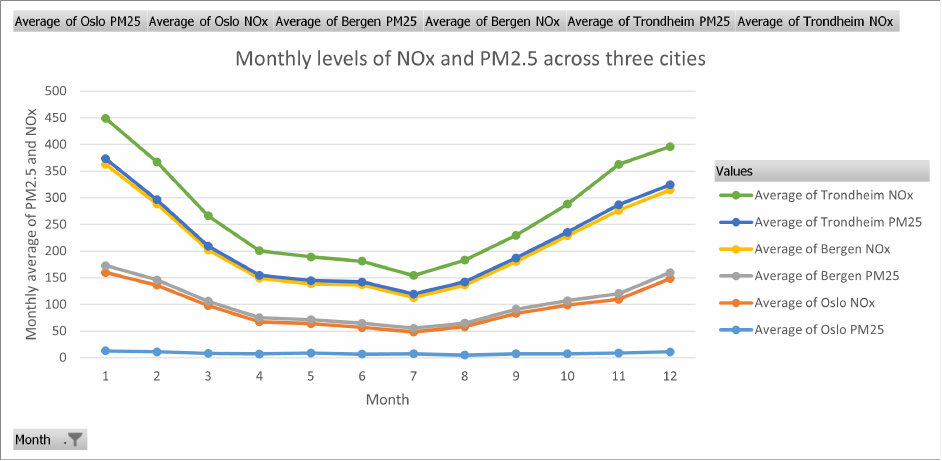}
    \caption{Monthly levels of two pollutants across three cities}
    \label{fig:Monthly}
\end{figure}

\par

\begin{figure}[H]
    \centering
    \includegraphics[width=\linewidth]{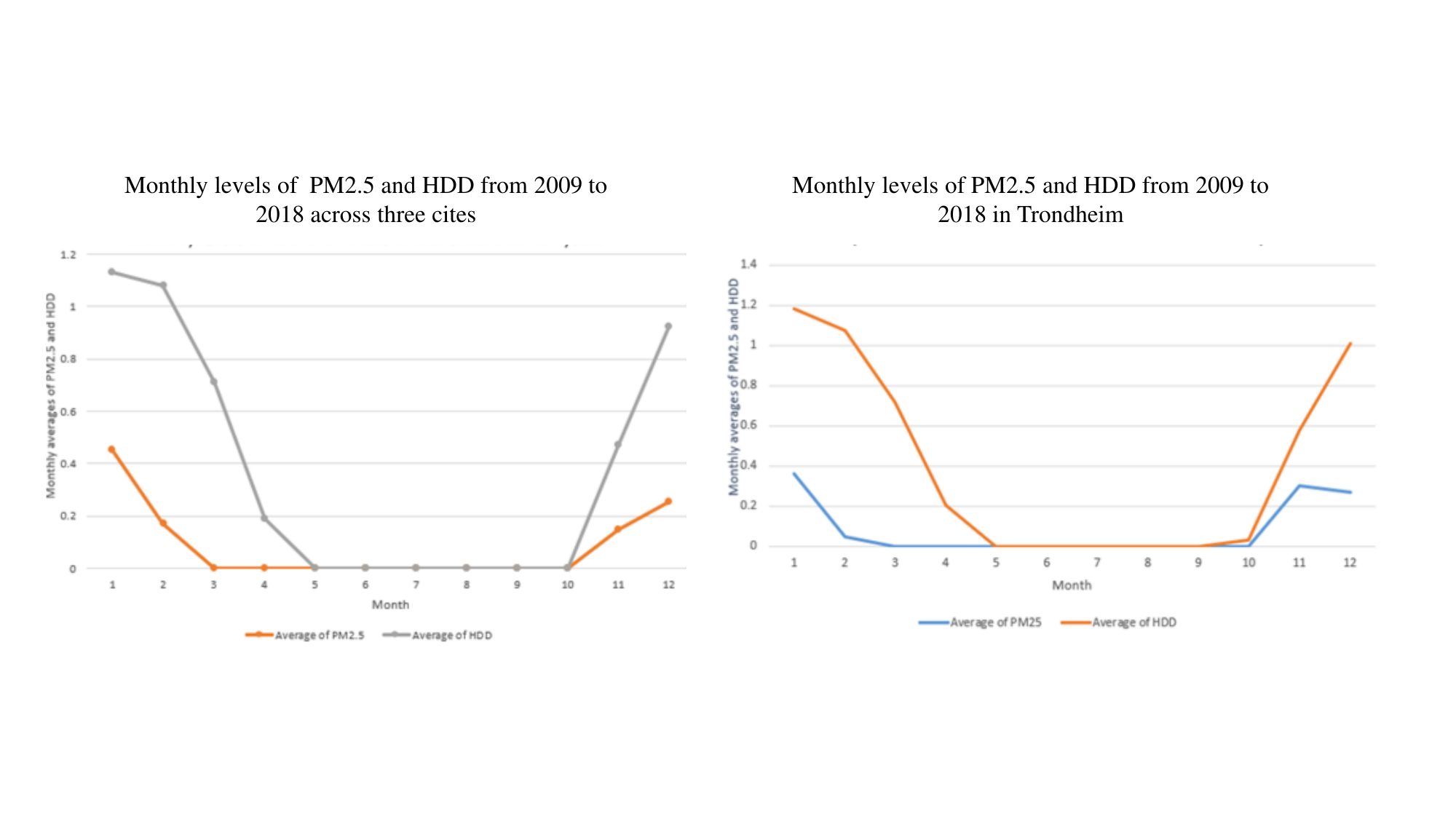}
    \caption{Monthly levels of two pollutants}
    \label{fig:Monthly}
\end{figure}

\par

  \begin{figure}[H]
    \centering
    \includegraphics[width=\linewidth]{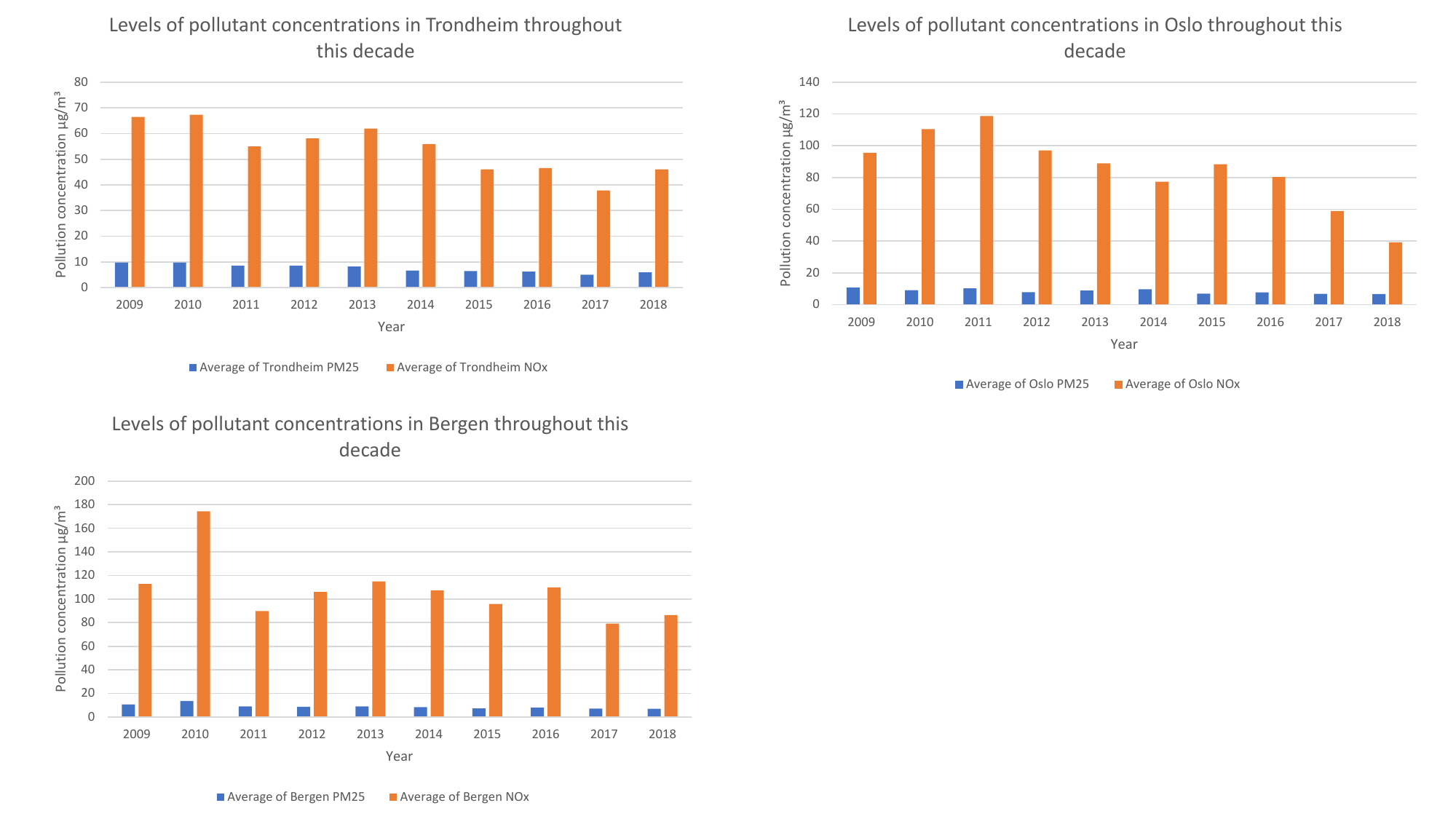}
    \caption{Yearly level of two pollutants}
    \label{fig:full_page_image}
  \end{figure}

\par

\section*{Appendix D Utilizing Python for Deep Learning Models Optimization and Evaluation with Hyperparameter Tuning}

We use Python, version 3.11.5 | packaged by Anaconda, Inc. | (main, Sep 11, 2023, 13:26:23) [MSC v.1916 64 bit (AMD64)]. Features and the target variable are extracted from the dataset, followed by standardization of the features. The dataset is then split into training and testing sets, with 80\% allocated for training, and 20\% reserved for testing. For LSTM modeling, a custom hypermodel class named MyHyperModel is defined, inheriting from Keras Tuner's HyperModel class. This class specifies the architecture of the LSTM model, including hyperparameters such as the number of LSTM units, dropout rates, and learning rates. In the context of Physical Based Deep Learning (PBDL), utilizing TensorFlow's Keras API, neural network models are established with three dense layers, each comprising 100 units. ELU activation functions and L2 regularization are employed to prevent overfitting, with batch normalization applied after each dense layer to stabilize training. Hyperparameter tuning is conducted to explore various hyperparameter combinations, aiming to minimize validation loss. To prevent overfitting, we integrate L2 regularization into each dense layer of both models. The best hyperparameters obtained from the tuning process are then used to construct the final model.\par

\begin{figure}[H]
    \centering
    \includegraphics[width=\linewidth]{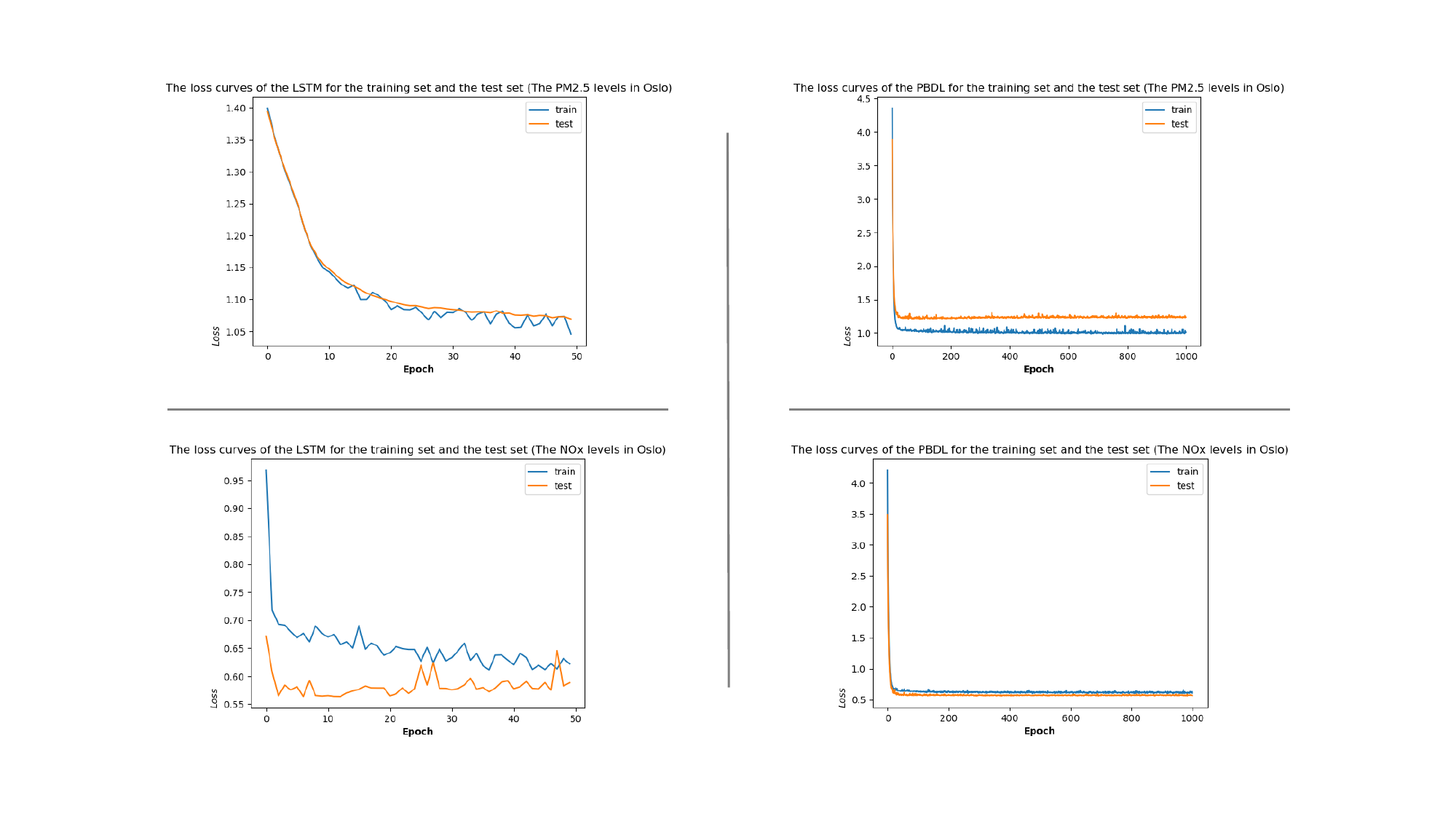} 
    \caption{The loss curve associated with Oslo's data}
    \label{fig:Oslo}
\end{figure}

\begin{figure}[H]
    \centering
    \includegraphics[width=\linewidth]{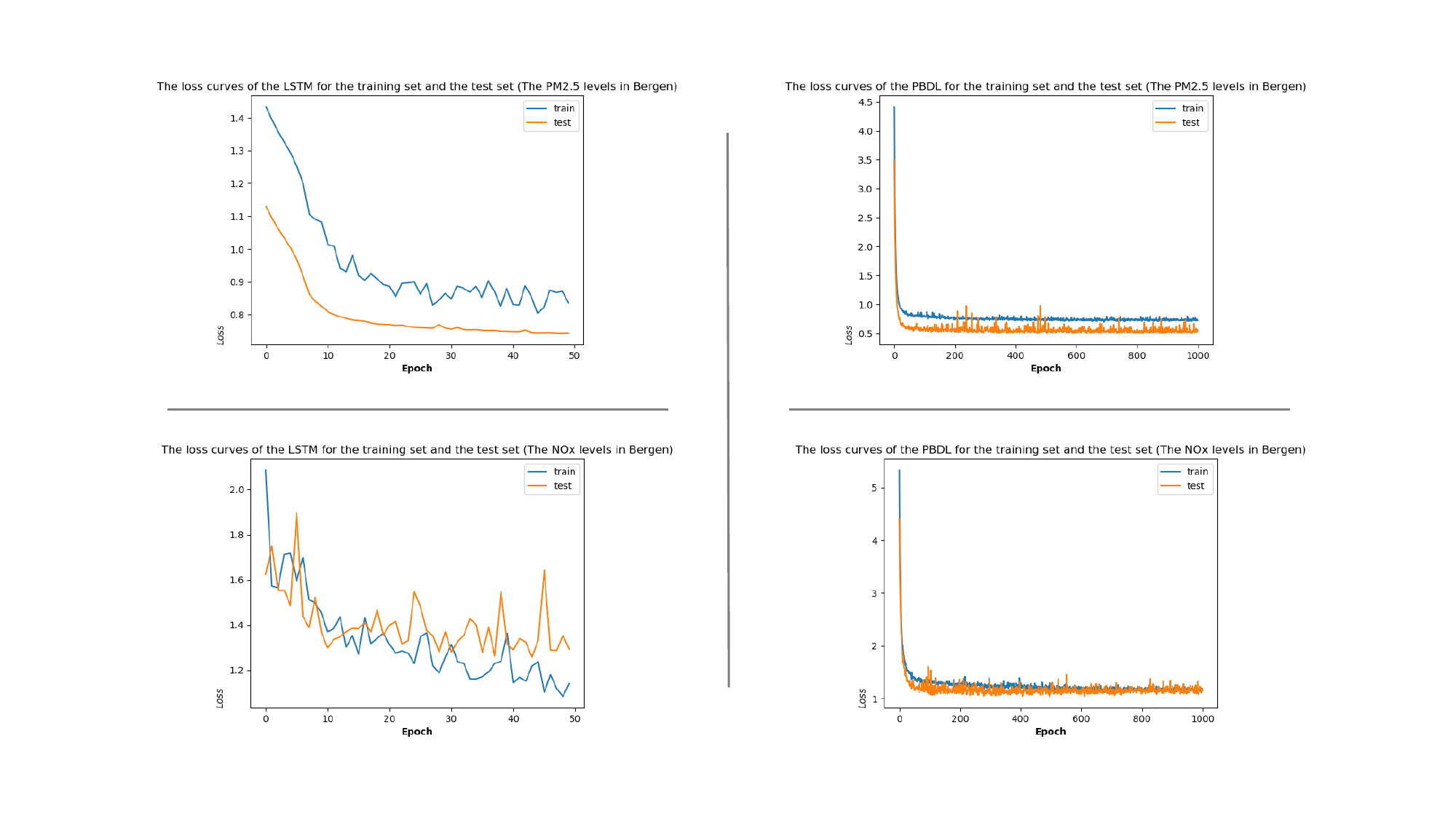} 
    \caption{The loss curve associated with Bergen's data }
    \label{fig:Bergen}
\end{figure}

\begin{figure}[H]
    \centering
    \includegraphics[width=\linewidth]{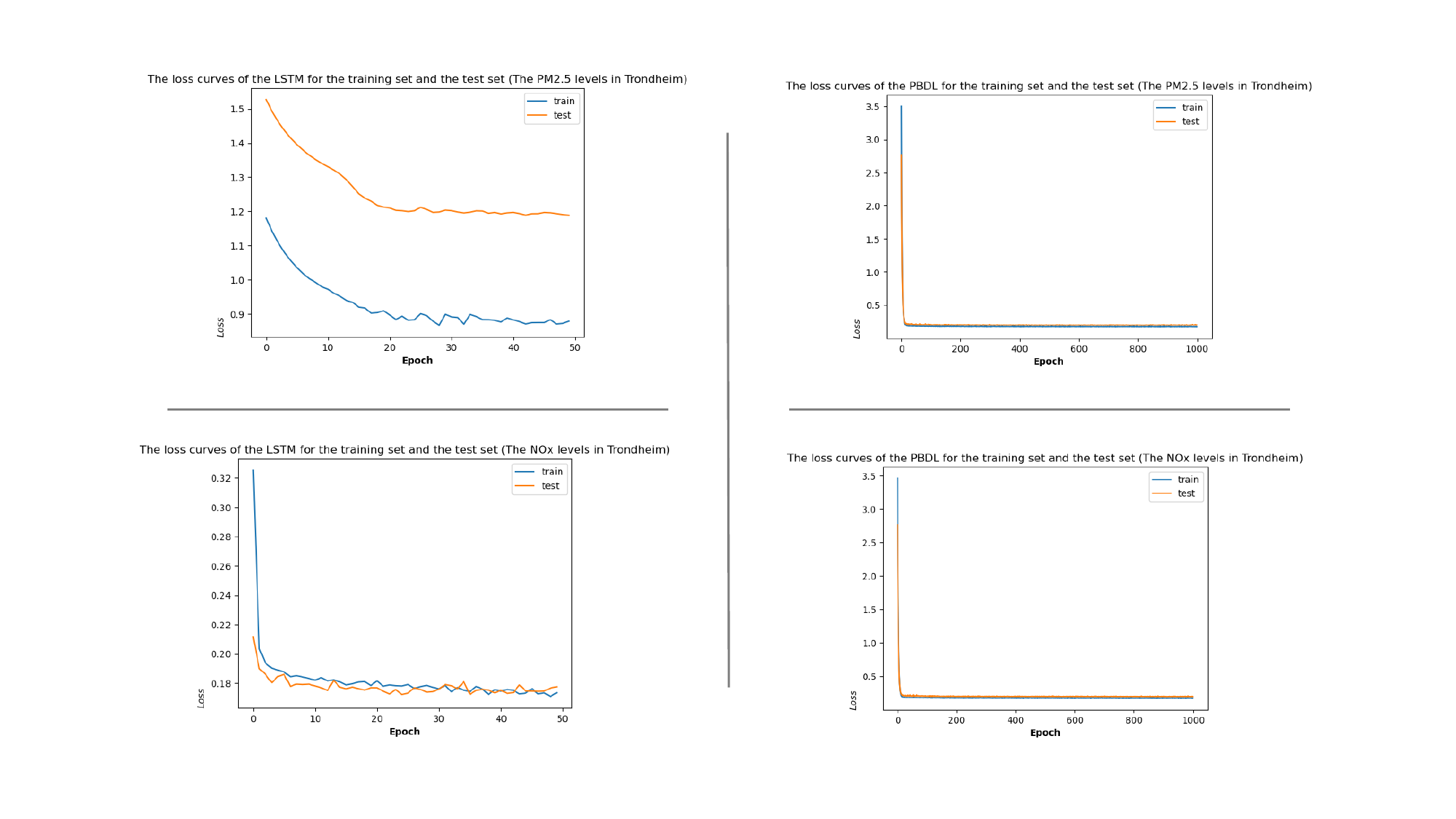} 
    \caption{The loss curve associated with Trondheim's data}
    \label{Trondheim}
\end{figure}

\begin{figure}[H]
    \centering
    \includegraphics[width=\linewidth]{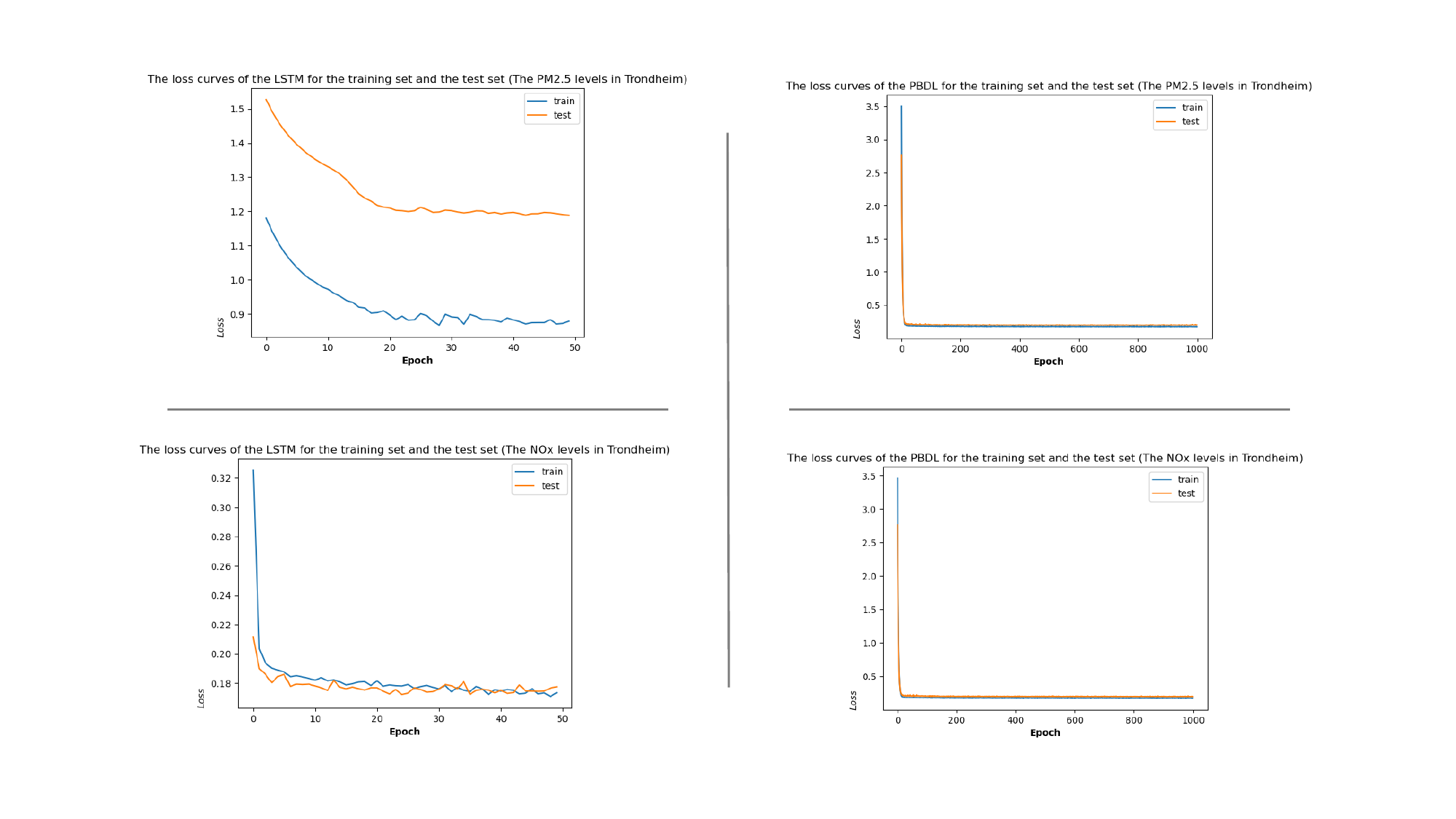} 
    \caption{Graphically comparing the predicted curve with the actual value curve for Oslo}
    \label{Oslo:e1}
\end{figure}

\begin{figure}[H]
    \centering
    \includegraphics[width=1\linewidth]{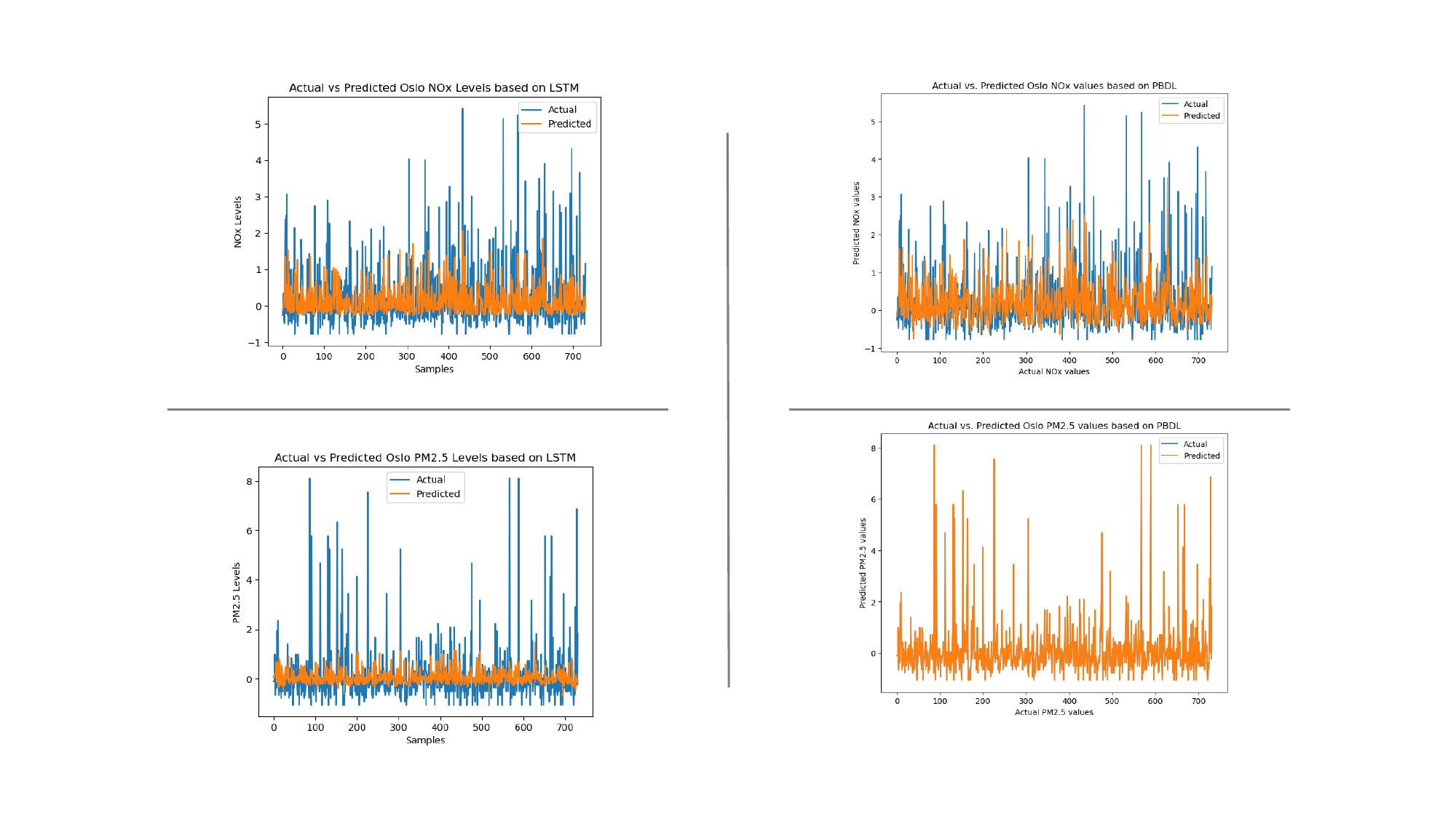} 
    \caption{Graphically comparing the predicted curve with the actual value curve for Bergen}
    \label{figBergenp2}
\end{figure}

\begin{figure}[H]
    \centering
    \includegraphics[width=1\linewidth]{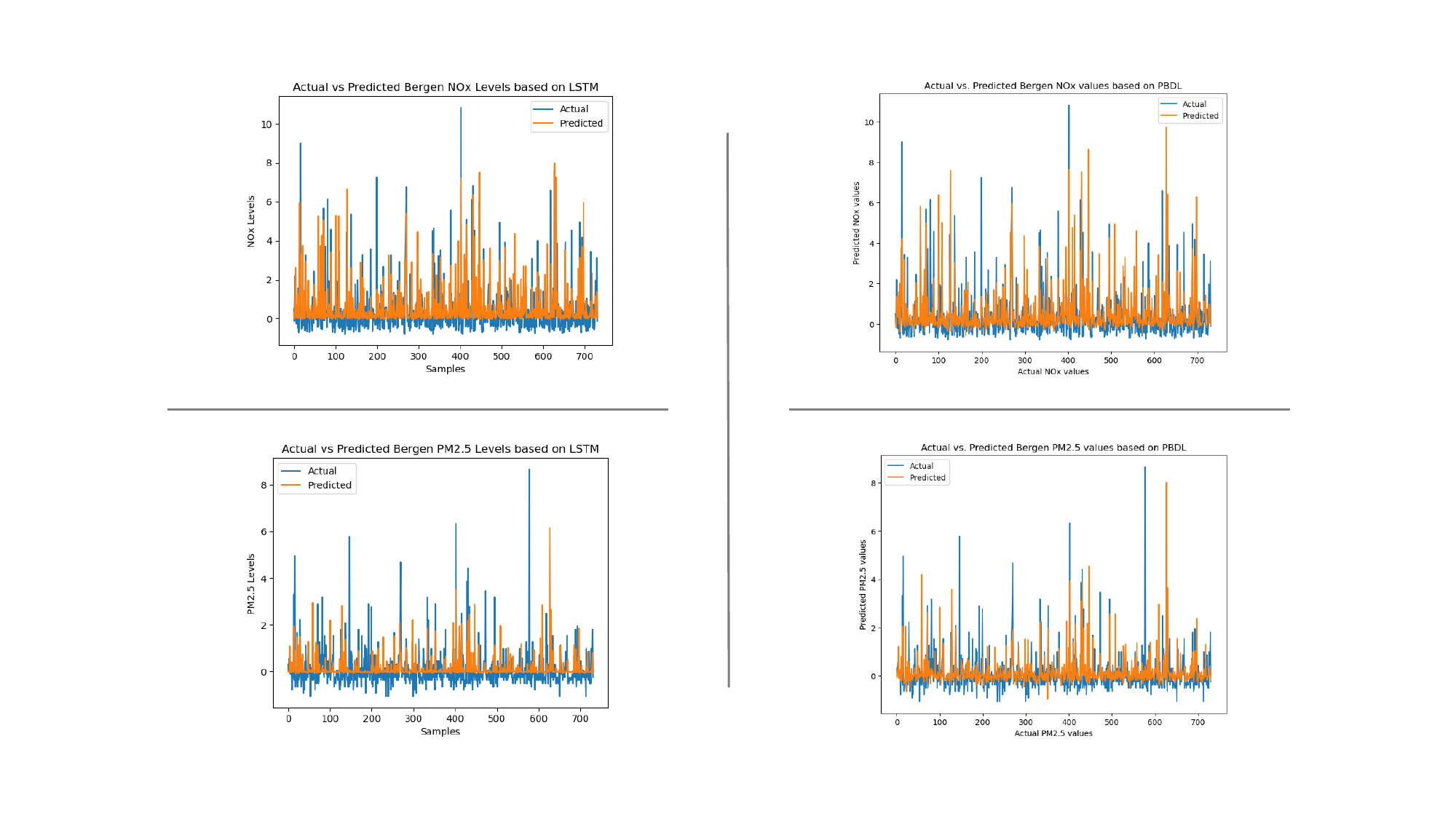} 
    \caption{Graphically comparing the predicted curve with the actual value curve for Trondheim}
    \label{fig:Trondheim}
\end{figure}

\begin{table}[H]
\centering
\caption{Hyperparameters Adjust Range for LSTM and PBDL Models}
\label{tab:hyperparameters_adjusted}
\begin{tabular}{@{}ll@{}}
\toprule
\textbf{Model}     & \textbf{Hyperparameters Adjusted}                                                                                         \\ \midrule
LSTM               & \begin{tabular}[c]{@{}l@{}}Units: 32, 48, 64, ..., 496, 512\\ Dropout Rate: 0.2, 0.4, 0.6, 0.8\\ LSTM Layers: 1, 2, 3, 4\\ Learning Rate: 1e-2, 1e-3, 1e-4\\ RandomSearch Trials: 10 with 2 runs each\end{tabular} \\[10pt]
PBDL               & \begin{tabular}[c]{@{}l@{}}Learning Rate: 0.0001, 0.0005, 0.001, 0.005, 0.01\\ Number of Units: 50 to 200\\ Number of Layers: 1 to 4\\ Kernel Regularizer: None, L2 (0.01), L2 (0.001)\end{tabular}          \\ \bottomrule
\end{tabular}
\end{table}

 \begin{table}[H]
\centering
\caption{Optimal Hyperparameters for LSTM}
\label{tab:lstm_hyperparameters}
\begin{tabular}{@{}lll@{}}
\toprule
\textbf{Parameter}     & \textbf{NOx}                  & \textbf{PM2.5}               \\ \midrule
units                  & 272                           & 320                           \\
dropout\_rate          & 0.4                           & 0.4                           \\
num\_layers            & 3                             & 1                             \\
units\_0               & 496                           & 272                           \\
dropout\_rate\_0       & 0.2                           & 0.4                           \\
units\_1               & 432                           & -                             \\
dropout\_rate\_1       & 0.6                           & -                             \\
units\_2               & 208                           & -                             \\
dropout\_rate\_2       & 0.2                           & -                             \\
units\_last            & 160                           & 48                            \\
dropout\_rate\_last    & 0.4                           & 0.8                           \\
learning\_rate         & 0.001                         & 0.0001                        \\
\bottomrule
\end{tabular}
\end{table}

\begin{table}[htbp]
\centering
\caption{Optimal  Hyperparameters for PBDL}
\label{tab:pbdl_hyperparameters}
\begin{tabular}{@{}ll@{}}
\toprule
\textbf{Hyperparameter}      & \textbf{Value}             \\ \midrule
Number of dense layers       & 1                          \\
Units per dense layer        & 107                        \\
Activation function          & ELU                        \\
Weight initialization        & He normal                  \\
Regularization               & L2 (0.01)                  \\
Batch Normalization          & Applied after each dense layer \\
Optimizer                    & Nadam                      \\
Learning rate                & 0.01                      \\
Number of epochs             & 1000                       \\
Batch size                   & 32                         \\ \bottomrule
\end{tabular}
\end{table}

\end{document}